\documentclass[journal]{IEEEtran}

\usepackage{cite}
\usepackage{amsmath,amssymb,amsfonts}
\usepackage{algorithmic}
\usepackage{graphicx}
\usepackage{textcomp}

\usepackage{algorithm}

\usepackage{verbatim}   
\usepackage{color}      
\usepackage{amsmath}    
\usepackage{amssymb}
\usepackage{amsmath}    
\usepackage{amsthm}     
\usepackage{subfigure}  
\usepackage{romannum}   
\usepackage{kotex}      
\usepackage{mathtools}  
\usepackage{makecell}   
\usepackage{multirow}
\usepackage{booktabs}   

\newcommand{\rom}[1]{\uppercase\expandafter{\romannumeral #1\relax}}  
\newcommand{\romlow}[1]{\lowercase\expandafter{\romannumeral #1\relax}}  
\newcommand{\our}{HGS}

\ifCLASSINFOpdf
\else
\fi

\begin{document}
%
\title{Learning-enabled Flexible Job-shop Scheduling for Scalable Smart Manufacturing}
%
%
%

\author{Sihoon Moon$^{1}$,
        Sanghoon Lee$^{1}$,
        and Kyung-Joon~Park$^{1, \dagger}$,~\IEEEmembership{Senior~Member,~IEEE}
        
\thanks{$^{1}$S. Moon, S. Lee, and K.-J. Park are with the Department of Electrical Engineering and Computer Science, DGIST, Daegu 42988, South Korea (e-mail: \{msh0576, leesh2913, kjp\}@dgist.ac.kr).}
\thanks{$\dagger$K.-J. Park is the corresponding author.}

}

\markboth{Journal of \LaTeX\ Class Files,~Vol.~14, No.~8, August~2015}%
{Shell \MakeLowercase{\textit{et al.}}: Bare Demo of IEEEtran.cls for IEEE Journals}

\maketitle

\begin{abstract}
In smart manufacturing systems (SMSs), flexible job-shop scheduling with transportation constraints (FJSPT) is essential to optimize solutions for maximizing productivity, considering production flexibility based on automated guided vehicles (AGVs). 
Recent developments in deep reinforcement learning (DRL)-based methods for FJSPT have encountered a \textit{scale generalization} challenge. 
These methods underperform when applied to environment at scales different from their training set, resulting in low-quality solutions. 
To address this, we introduce a novel graph-based DRL method, named the Heterogeneous Graph Scheduler (\our{}). 
Our method leverages locally extracted relational knowledge among operations, machines, and vehicle nodes for scheduling, with a graph-structured decision-making framework that reduces encoding complexity and enhances scale generalization.
Our performance evaluation, conducted with benchmark datasets, reveals that the proposed method outperforms traditional dispatching rules, meta-heuristics, and existing DRL-based approaches in terms of makespan performance, even on large-scale instances that have not been experienced during training.

\end{abstract}


%
\IEEEpeerreviewmaketitle

\section{Introduction}
\label{sec:introduction}
As the field of smart manufacturing continues to evolve, a burgeoning array of novel information technologies, including the Internet of things (IoT), cloud computing, big data, and artificial intelligence, are increasingly being integrated into manufacturing processes to enhance production efficiency and flexibility~\cite{b17, b19, 
b38, b40}.
Recently, numerous enterprises in real-world manufacturing have employed transportation resources such as automated guided vehicles (AGVs) to improve flexibility and diversity in flexible manufacturing systems~(FMS)~\cite{b18}.
This can be mathematically formulated as a flexible job-shop scheduling problem with transportation constraints (FJSPT)~\cite{b9}.
However, due to the increased complexity of production scheduling, this problem poses significant challenges such as the allocation of operations to compatible machines and the assignment of AGVs for conveying intermediate products.
Deep reinforcement learning (DRL)-based FJSPT schedulers have emerged as a promising approach, offering the potential to discover near-optimal solutions with reduced computation time~\cite{b10, b11, b39}.

The main challenge of this study is the issue of \textit{scale generalization}.
In real-world scenarios, smart manufacturing systems frequently encounter alterations in the process environment (i.e., scale changes), such as the insertion of new jobs or the addition/breakdown of machines and vehicles~\cite{b17}.
Conventional DRL-based schedulers are trained on specific-scale instances, characterized by a fixed number of operations, machines, and vehicles. While these schedulers perform effectively on the trained instances and similar-sized unseen instances, their efficacy diminishes substantially for unseen large-scale instances~\cite{b32, 
b22
}.
This means that as the manufacturing environment changes, the scheduler may produce low-quality solutions, resulting in a loss of productivity.
Furthermore, it is impractical and costly to constantly retrain the scheduler in response to scale changes.
Therefore, it is necessary to design a scale-agnostic DRL-based FJSPT scheduler that can provide a near-optimal solution even for unseen large-scale instances.

There is also a technical challenge of DRL-based FJSPT scheduler; \textit{end-to-end decision making}.
Numerous DRL-based methods for manufacturing process scheduling adopt a rule-based decision-making framework~\cite{b31, b36, b41}, which selects one of the predefined dispatching rules at the decision-time step.
However, this approach has the disadvantage of heavily relying on expert experience due to the design of the rules, and it lacks sufficient exploration of the action space~\cite{b34, b5}.
Especially in FJSPT, it is difficult to anticipate sufficient action space exploration, as it requires not only the selection of operations and machines, but also the selection of vehicles.
Therefore, it is necessary for the scheduler to be able to select the most valuable set of operations, machines, and vehicles at each decision time step to minimize makespan.

To address these challenges, we propose a graph-based DRL module for solving FJSPT, called Heterogeneous Graph Scheduler (\our{}), which consists of three main components; a \textit{heterogeneous graph structure}, a \textit{structure-aware heterogeneous graph encoder}, and a \textit{three-stage decoder}.
To address the scale generalization challenge, we develop a novel heterogeneous graph structure capable of representing FJSPT and a structure-aware heterogeneous graph encoder.
The design of this graph structure incorporates operation, machine, and vehicle nodes, all interconnected through edges, which symbolize processing and transportation times.
The encoder consists of three sub-encoders (operation, machine and vehicle sub-encoders) and a global encoder.
The intuitive idea behind scale generalization is that relationships between adjacent nodes at a small-scale instance will be effective in solving large-scale problems. 
To learn this relationality well, we perform local encoding per node type, preferentially incorporating information from highly relevant neighbors. Then, we encode the entire graph using global encoding.
For example, in the proposed graph structure, machine nodes are directly adjacent exclusively to operation nodes, while vehicle nodes maintain their adjacency solely to operation nodes.
From the perspective of machine nodes, it is preferable to consider the processing time of operations rather than the transportation time of vehicles.
The sub-encoder of a machine node can encode only the processing time, not including transportation time, thus reducing the complexity of the encoding. 
Because this method partitions and encodes information locally, it can effectively integrate knowledge even when the environment changes in large-scale instances.
Therefore, this structure-aware encoder significantly improves the scale generalization capabilities, particularly in unseen large-scale instances.

In an attempt to tackle the end-to-end decision-making challenge, we propose the development of a three-stage decoder model.
The decoder adopts an end-to-end decision-making framework that directly outputs scheduling solutions.
At each decision-time step, the decoder utilizes the graph embedding knowledge derived from the encoder. This guides the decision on the assignment of operations to machines, as well as the allocation of vehicles for transportation (operation-machine-vehicle pair).
In the initial stage, the decoder selects an operation node that is most relevant to the context node, which incorporates both the graph embedding knowledge from the encoder and the knowledge from previous actions.
In the second stage, given the already selected operation embedding, the decoder proceeds to select the machine node that is most relevant for the context node. 
In the final stage, it does the same for the vehicle node.
This means that the decoder sequentially selects the nodes from each class that are most likely to minimize the makespan based on the context nodes.

Our contributions can be summarized as follows:
\begin{itemize}
    \item We develop a novel heterogeneous graph structure tailored for FJSPT and a structure-aware heterogeneous graph encoder to represent the proposed heterogeneous graph.
    This encoder enables each node to structurally aggregate messages from neighboring nodes of diverse classes, such as operations, machines, and vehicles.  
    \item We construct a three-stage decoder specifically customized for FJSPT. 
    Through the three-stage decoding, the decoder sequentially selects operation, machine and vehicle nodes based on the graph embedding derived from the encoder at every decision-time step.
    The composite action produces high-quality FJSPT scheduling solutions to minimize the makespan.
    \item By integrating the encoder, decoder, and RL framework, we develop the \our{} module.
    The proposed method outperforms traditional dispatching rules, meta-heuristic methods, and existing DRL-based algorithms, especially for the scale generalization capabilities. 
    Moreover, we validate the superiority of the proposed method by conducting simulations on a variety of benchmark datasets~\cite{b25}.
    To the best of our knowledge, the proposed method is the first approach for the size-agnostic DRL-based FJSPT scheduler.
\end{itemize}

\section{Related work}

\subsection{Conventional scheduling methods for FJSPT}
Conventional methods to solve FJSPT can be classified into exact methods, heuristics and meta-heuristics.
Exact methods, such as mixed-integer linear programming~\cite{b25} and constraint programming~\cite{b27}, are capable of finding the optimal solution, but at the expense of an exponentially high computational cost. 
Meta-heuristic methods are advanced search strategies to find high-quality solutions in a reasonable time.
For instance, the study in~\cite{b30}, develops a genetic algorithm (GA) specifically for multi-objective optimization that addresses both makespan and energy consumption minimization simultaneously. 
The work presented in~\cite{b12} introduces an ant colony optimization (ACO) algorithm that accounts for sequence-dependent setup time constraints within the FJSPT.
In the paper~\cite{b9}, a particle swarm optimization (PSO) algorithm, augmented with genetic operations, is developed to solve dynamic FJSPT (DFJSPT), considering dynamic events such as the arrival of new jobs, machine breakdowns, and vehicle breakdowns/recharging.
While these methods can find near-optimal solutions, they are computationally inefficient because they require extensive exploration in a large search space.
This restricts their practical applicability in rapidly changing environments, where unforeseen disturbances can occur even before a revised rescheduling plan is developed.

\subsection{DRL-based scheduling methods for FJSPT}
In the recent years, an increasing number of researchers have been applying DRL techniques to complex scheduling problems such as FJSP and have achieved remarkable results.
The authors in~\cite{b31} utilize DRL to jointly optimize makespan and energy consumption in FJSPT, while also addressing dynamic environment issues such as new job insertion.
However, this method defines the action space of the DRL model as a combination of dispatch rules. Using the composite dispatching rules as  actions, instead of directly finding scheduling solutions, relies heavily on the quality of the rules and human experiences.
Conversely, study~\cite{b32} introduces an end-to-end DRL framework for FJSPT, wherein a DRL agent, at every decision step, determines the vehicle that should transport a specific job to a particular machine.
Their model, however, is demonstrated only for small-scale instances, with no consideration for large-scale generalization.
Several studies investigating scale generalization~\cite{b10, b34} have been conducted in FJSP. To accommodate this characteristic, they implement a GNN-based DRL framework. 
Existing DRL models, such as multi-layer perceptron (MLP) or convolutional neural network (CNN), use vectors or matrices to represent states.
The main drawback of these representations is that the vector size is fixed, making them inflexible and unable to solve problems of varying sizes.
GNN, on the other hand, can handle graphs of varying sizes, overcoming the limitations of vector representations~\cite{b14}.
Nevertheless, these studies focus only on FJSP without considering transportation resources.

\section{Preliminary}

\subsection{Problem description and notations}
FJSPT can be defined as follows.
There exists a set of $n$ jobs $\mathcal{J}=\{J_1,...,J_n \}$, a set of $m$ machines $\mathcal{M}= \{ M_1,..., M_m\}$ and a set of $v$ vehicles $\mathcal{V} = \{ V_1,..., V_v\}$.
Each job $J_i$ consists of $n_i$ consecutive operations $\mathcal{O}_i = \{ O_{i1}, ..., O_{in_i} \}$ with precedence constraints.
An operation of job $J_i$ denoted by $O_{ij}$ can be processed on a subset of eligible machines $\mathcal{M}_{ij} \subset \mathcal{M}$.
This implies that operation $O_{ij}$ requires different processing times $T_{ijk}^p$ for each machine $M_k \in \mathcal{M}_{ij}$.
To allocate operation $O_{ij}$ on machine $M_k$, vehicle $V_u$ transports the intermediate products of $O_{ij}$ to the machine.
Transportation time of $V_u$ is the sum of off-load and on-load transportation time.
The off-load time $T_{iju}^t$ indicates that $V_u$ approaches the product location of $O_{ij}$ to load it in an off-load status.
The on-load time $T_{kk'}^t$ indicates that the vehicle transports the product from machine $M_{k}$ to machine $M_{k'}$ in an on-load status, where we assume that the on-load transportation time between two machines is the same for all vehicles.
In FJSPT, the optimization goal is to assign operations to compatible machines and select vehicles to transport them, while determining the sequence of operation-machine-vehicle pairs to minimize the makespan,
\begin{equation}
\begin{aligned}
    C_{max} = \max C_{in_i}, \forall i \in \{ 1,...,n \},
\end{aligned}
\end{equation}
where $C_{in_i}$ is a completion time of final operation $O_{in_i}$ of job $J_i$.

\begin{table}[]
\begin{tabular}{ll}

\hline
\textbf{Constants} &                                                \\
$m$         & total number of machines                       \\
$n$         & total number of jobs                           \\
$n_i$      & total number of operations of job $i$            \\
$v$         & total number of vehicles                      \\
$d_h$       & dimension of node embedding    \\
$d_e$       & dimension of edge embedding    \\
$d_\text{k}$       & dimension of query and key \\
$d_\text{v}$        & dimension of value\\
\textbf{Indexes}   &                                                \\
$i$         & job index, $i=1,...,n$                           \\
$j$         & operation index of $J_i, j=1,..., n_i$ \\
$k$         & machine index, $k=1,...,m$             \\
$u$         & vehicle index, $u=1,...,v$             \\
\textbf{Sets}        &   \\
$\mathcal{J}$    & job node set, $\mathcal{J}=\{ J_1, ..., J_n  \}$   \\
$\mathcal{O}_i$     & operation node set for job $J_i$, $\mathcal{O}_i= \{ O_{i1}, ..., O_{in_i} \}$    \\
$\mathcal{M}$    & machine node set, $\mathcal{M} = \{ M_1, ..., M_m \}$ \\
$\mathcal{M}_{ij}$  & available machine node set for operation $O_{ij}$, $\mathcal{M}_{ij} \subseteq \mathcal{M}$   \\
$\mathcal{V}$   & vehicle node set, $\mathcal{V} = \{ V_1, ..., V_v \}$  \\
$\mathcal{N}_{m}(O_{ij})$  & neighboring machine nodes for $O_{ij}$ \\
$\mathcal{N}_{v}(O_{ij})$  & neighboring vehicle nodes for $O_{ij}$   \\
$\mathcal{N}(M_{k})$        & neighboring operation nodes for $M_k$ \\
$\mathcal{N}(V_{u})$        & neighboring operation nodes for $V_u$ \\
\textbf{Variables}   &   \\
$T_{ijk}^{p}$   & processing time of operation $O_{ij}$ on machine $M_k$    \\
$T_{u}^{t}$    & transportation time of vehicle $V_u$ \\
$T_{kk'}^t$    & time that a vehicle transports products from $M_k$ to $M_{k'}$  \\
$T_{ij}^s$  & start time of $O_{ij}$ \\
$C_i$       & completion time of job $J_i$  \\

\hline

\end{tabular}
\caption{Notation.}
\vspace{-20pt}
\label{tab:notation}
\end{table}

\subsection{Disjunctive graph for FJSP}
\begin{figure}[t]
    \centering
    \subfigure[disjunctive graph example.]{
    \includegraphics[width=0.45 \linewidth]{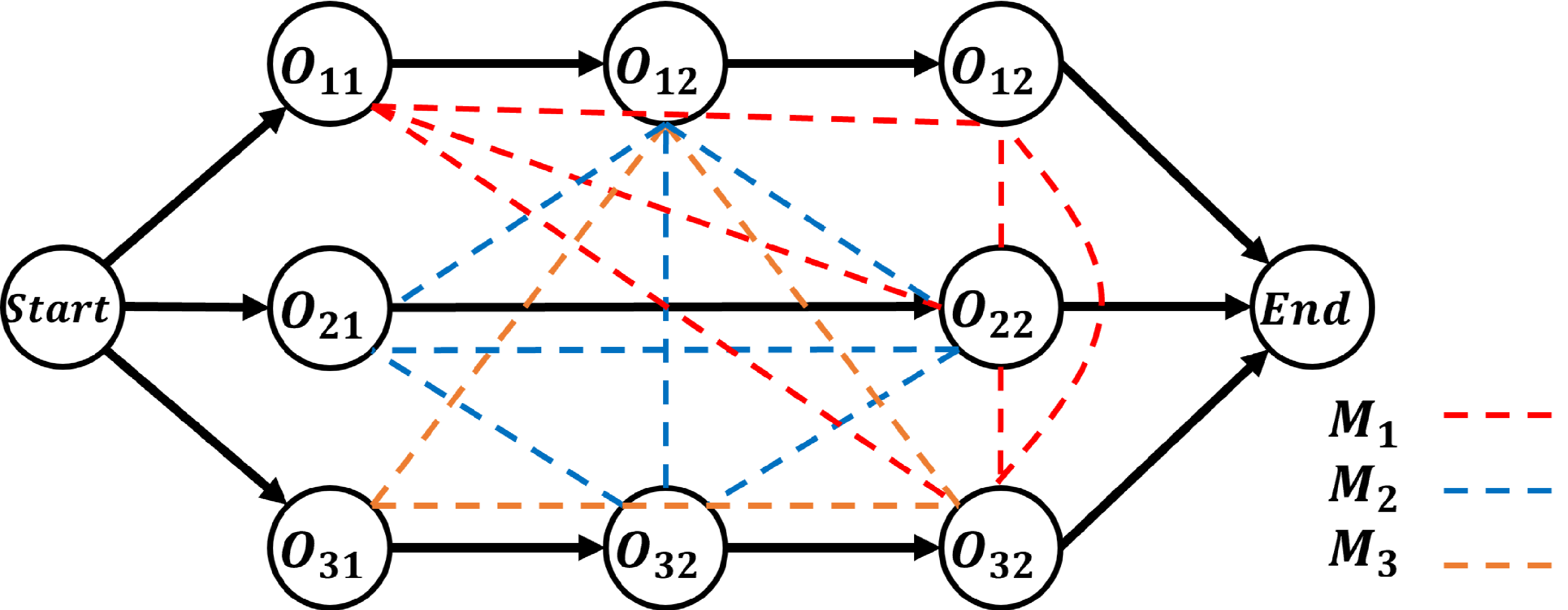}
    }
    \subfigure[disjunctive graph solutions.]{
    \includegraphics[width=0.45 \linewidth]{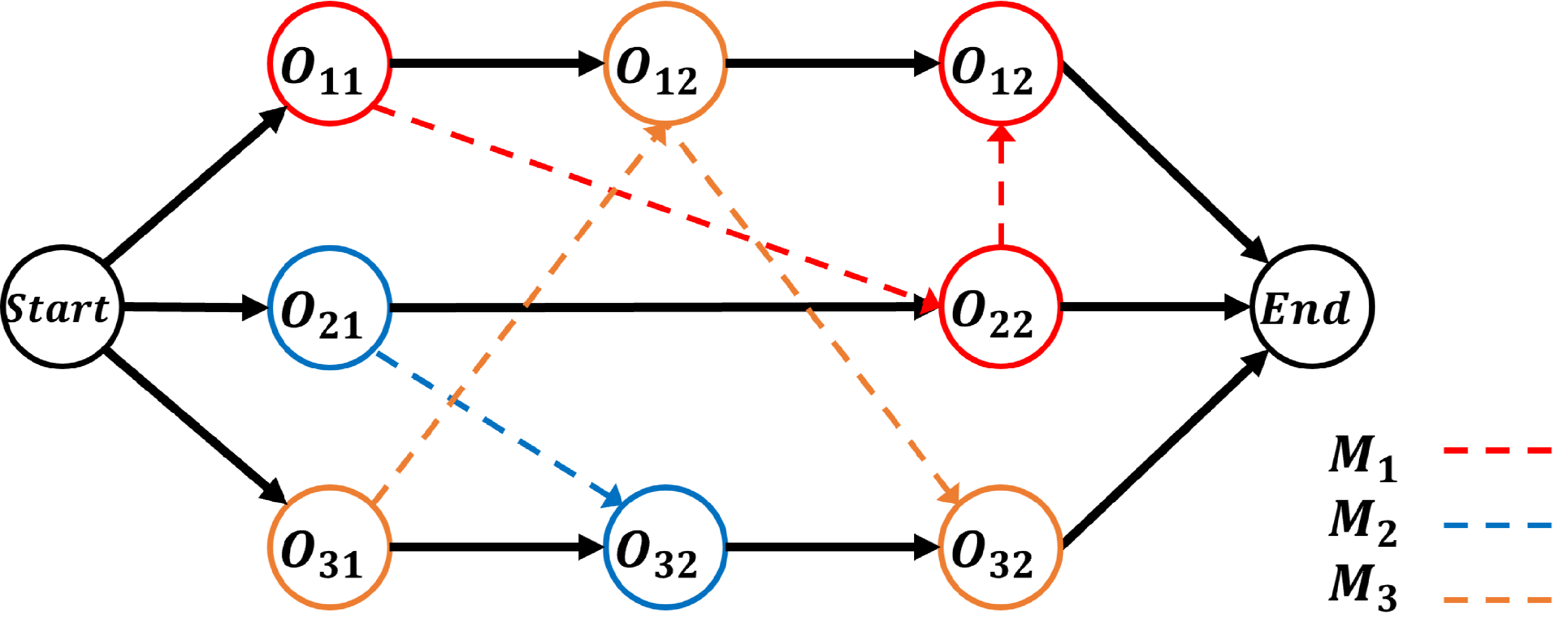}
    \label{fig:disjunctive_graph_solution}
    }
    \caption{Disjunctive graph for FJSP.}
    \label{fig:disjunctive_graph}
    \vspace{-10pt}
\end{figure}

Fig.~\ref{fig:disjunctive_graph} presents a disjunctive graph for FJSP~\cite{b4}, which is denoted as $\mathcal{G=(O, C, D)}$.
$\mathcal{O}=\{ O_{ij} | \forall i, j \} \cup \{ Start, End \}$ comprises the set of operation nodes, including all operations and two dummy nodes (with zero processing time) that indicate the start and end of production.
$\mathcal{C}$ is the set of conjunctive arcs.
There are $n$ directed arc flows connecting the $Start$ and $End$ nodes, with each of the flows illustrating the processing sequence for job $J_i$.
$\mathcal{D}=\cup_k \mathcal{D}_k$ constitutes a set of (undirected) disjunctive arcs, where $\mathcal{D}_k$ forms a clique that links operations capable of being executed on machine $M_k$.
Given that operations in FJSP can be performed on multiple machines, an operation node can be linked to multiple disjunctive arcs.
Solving FJSP involves selecting a disjunctive arc for each node and establishing its direction, as illustrated in Fig.~\ref{fig:disjunctive_graph_solution}.
The directed arcs imply a processing sequence of operations on a machine.

\subsection{Attention model}
The Attention Model (AM)~\cite{b8} is a weighted message-passing technique between nodes in a graph, and it learns attention scores between nodes based on how much they relate to each other.
In the graph, there is only one class of node $x\in X$.
Let $h_x \in \mathbb{R}^{d_h}$ represent the embedding vector of node $x$, with $d_h$ being the embedding dimension.
The model necessitates three vectors - \textit{query} $\text{q}$, \textit{key} $\text{k}$, and \textit{value} $\text{v}$ - to form the aggregated node embedding.
These are expressed as follows, 
\begin{equation}
\begin{aligned}
\label{eqn:qkv}
    \text{q}_{x}=W^{\text{q}}h_{x}, 
    \text{k}_y=W^{\text{k}}h_y,
    \text{v}_y=W^{\text{v}}h_y,
    \; x, y \in X
\end{aligned}
\end{equation}
where $W^{\text{q}}, W^{\text{k}}\in \mathbb{R}^{d_\text{k} \times d_h}$ and $W^{\text{v}}\in \mathbb{R}^{d_\text{v} \times d_h}$ are the trainable parameter matrices. 
$d_\text{k}$ is the query/key dimensionality, and $d_\text{v}$ is the value dimensionality. 
In cases where $y=x$, this mechanism is referred to as self-attention.
Utilizing the query $\text{q}_x$ from node $x$ and the key $\text{k}_y$ from node $y$, the \textit{compatibility} $\sigma_{xy}$ is determined through the scaled dot-product:
\begin{equation}
\begin{aligned}
\label{eqn:compatability}
    \sigma_{xy}=
    \begin{cases}
    \frac{\text{q}_x^T \text{k}_y}{\sqrt{d_k}} & \text{if }y \text{ is a neighbor of }x \\
    -\infty & \text{otherwise}
    \end{cases}
\end{aligned}
\end{equation}
where $-\infty$ prevents message passing between non-adjacent nodes. From the compatibility, we compute \textit{attention weights} $\Bar{\sigma}_{xy}\in[0,1]$ using a softmax: 
\begin{equation}
\begin{aligned}
\label{eqn:score}
    \Bar{\sigma}_{xy}=\frac{e^{\sigma_{xy}}}{\sum_{y'\in X} e^{\sigma_{xy'}}}.
\end{aligned}
\end{equation}
This measures the importance between $x$ and $y$; a greater attention weights $\Bar{\sigma}_{xy}$ implies a higher dependence of node $x$ on node $y$. Subsequently, the attention-based single-head node embedding $h_{x,z}'$ for node $x$ is calculated as a weighted sum of messages $\text{v}_y$: 
\begin{equation}
\begin{aligned}
\label{eqn:single-head attention}
    h_{x,z}'=\sum_{y\in X}\bar{\sigma}_{xy} \text{v}_y,
\end{aligned}
\end{equation}
where $z \in \{1,..., Z\}$ is a head index.

The multi-head attention (MHA) enables a node to obtain neighboring messages from various attention types, executed $Z$ times in parallel $(Z=8)$, with $d_{\text{k}} = d_{\text{v}} = \frac{d_h}{Z}$. 
The ultimate multi-head attention value for node $x$ is determined by summing the heads from all attention types.
This can be expressed as a function of node embeddings for all nodes:
\begin{equation}
\begin{aligned}
\label{eqn:multi-head attention}
    h_x = & \text{MHA}_x(\{ h_y|y\in X \}) \\
    = & \sum_{z\in Z} W_{x,z} h_{x,z}'  ,
\end{aligned}
\end{equation}
where $W_{x,z} \in \mathbb{R}^{d_h \times d_{\text{v}} }$ is a trainable parameter matrix.

\section{Heterogeneous graph scheduler (\our{})}
\begin{figure}[t]
    \centering
    \includegraphics[width=1.0\linewidth, height=5cm]{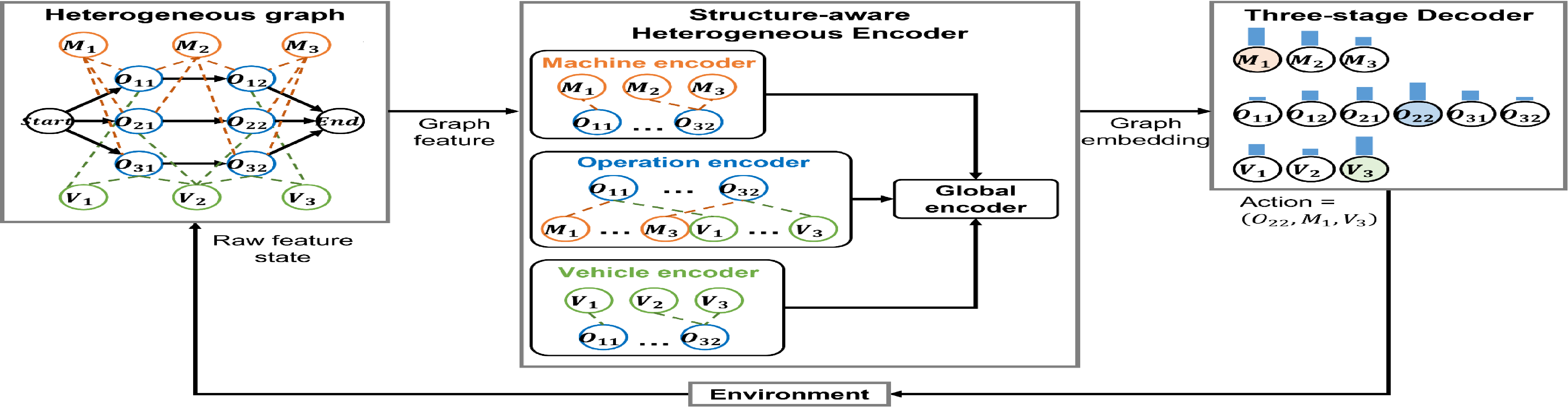}
    \caption{Heterogeneous graph scheduler architecture.}
    \label{fig:proposed_method_archit}
    \vspace{-10pt}
\end{figure}

To resolve FJSPT, we propose \our{} module consisting of three main components: a heterogeneous graph, a structure-aware heterogeneous encoder and a three-stage decoder. 
The workflow of the \our{} module, as depicted in Fig.~\ref{fig:proposed_method_archit}, unfolds in the following sequence: 1) receiving raw feature states from the manufacturing environment, 2) constructing the heterogeneous graph based on these features, 3) encoding the graph using the encoder, 4) determining a composite action involving an operation-machine-vehicle pair utilizing the decoder, and 5) repeating this process until all operations have been scheduled.
Initially, we develop a heterogeneous graph specifically tailored for FJSPT. This graph effectively encapsulates the features of operations, machines, vehicles, and their interrelationships, while maintaining a low graph density.
Next, we represent this heterogeneous graph using our proposed encoder, which incorporates three sub-encoders and a global encoder. 
Each sub-encoder enables a specific node to locally aggregate messages from adjacent nodes belonging to different classes. 
Subsequently, the global encoder integrates the encoded messages from all nodes.
Based on the graph representation, the decoder generates a composite action of the operation-machine-vehicle (O-M-V) pairs at the decision time step. 
Finally, using an end-to-end RL algorithm, we train the \our{} module to minimize the makespan.

\subsection{Heterogeneous graph for FJSPT}
The traditional disjunctive graph in Fig.~\ref{fig:disjunctive_graph} is difficult to represent FJSPT.
This is because, first, it does not include vehicle properties, such as the number of vehicles, their location, transportation time, and status (on-load or off-load).
Second, the disjunctive arc set $\mathcal{D}$ becomes much larger as the graph size (the number of nodes) increases.
The high-density graph leads to limited graph neural network performance~\cite{b5}.
Lastly, the traditional graph struggles to represent the processing time for compatible machines of an operation.
Thus, to resolve these issues, we propose a novel heterogeneous graph $\mathcal{H}$ for FJSPT.

By modifying the disjunctive graph, we propose a novel heterogeneous graph to represent FJSPT, as shown in FIg.~\ref{fig:proposed_method_archit}.
The graph is defined as $\mathcal{H}=(\mathcal{O} \cup \mathcal{M} \cup \mathcal{V}, \mathcal{C}, \mathcal{E}_m \cup \mathcal{E}_v^{\text{off}} \cup \mathcal{E}_v^\text{on})$.
We model the manufacturing environment as the graph. For example, we represent the processing time that an operation can be processed on availabe machines, the time it takes for a designated vehicle to load the product at the location of an finished operation and the time it takes to transport it to the next machine.
In contrast to the traditional graph, the machine node set $\mathcal{M}$, vehicle node set $\mathcal{V}$, compatible machine arc set $\mathcal{E}_m$ and compatible vehicle arc set $\mathcal{E}_v$ are added.
Machine node $M_k\in \mathcal{M}$ and vehicle node $V_u \in \mathcal{V}$ represent machine and vehicle features, respectively.
The disjunctive arc set $\mathcal{D}$ is replaced with $\mathcal{E}_m \cup \mathcal{E}_v^{\text{off}} \cup \mathcal{E}_v^\text{on}$.
An element of the compatible machine arc set $E_{ijk}^m \in \mathcal{E}_m$ denotes the processing time when operation $O_{ij}$ is processed on compatible machine $M_k \in \mathcal{M}_{ij}$.
Vehicle arc $E_{iju}^v \in \mathcal{E}_v^{\text{off}}$ represents the off-load transportation time for $V_u$ to arrive at the location of the product involved in $O_{ij}$, and arc $E_{kk'}^v \in \mathcal{E}_v^\text{on}$ represents on-load transportation time during which a vehicle in an on-load status moves from $M_k$ to $M_{k'}$.

In FJSPT, the heterogeneous graph has dynamic structure, $\mathcal{H}_t(\mathcal{O} \cup \mathcal{M} \cup \mathcal{V}, \mathcal{C}, \mathcal{E}_{mt} \cup \mathcal{E}_{vt}^\text{off} \cup \mathcal{E}_{v}^\text{on})$, where $\mathcal{E}_{mt}$ and $\mathcal{E}_{vt}^\text{off}$ are dynamically change during resolving FJSPT.
At time step $t$, once an action $(O_{ij}, M_{k}, V_{u})$ is selected, $\mathcal{H}_t$ transits to $\mathcal{H}_{t+1}$.
At time step $t$, the DRL model selects an action $(O_{ij}, M_{k}, V_{u})$, indicating that operation $O_{ij}$ is designated to be processed on machine $M_k$ and transported using vehicle $V_u$. 
Consequently, upon action selection, the state $\mathcal{H}_t$ transitions to $\mathcal{H}_{t+1}$, wherein only the selected edges $E_{ijk}^m \in \mathcal{E}_{mt}$ between $O_{ij}$ and $M_k$, and $E_{iju}^v \in \mathcal{E}_{vt}^\text{off}$ between $O_{ij}$ and $V_u$ are retained, while other edges compatible with the machine and vehicle are eliminated. 
Additionally, due to the selection of machine $M_k$ and vehicle $V_u$, edges $E_{i'j'k}$ and $E_{i'j'u}$ corresponding to another operation $O_{i'j'}$ are removed at time $t+1$ owing to the preemption of $M_k$ and $V_u$. 
This is under the premise that $O_{i'j'}$ has them listed within its compatible machine and vehicle sets, $M_k \in \mathcal{M}_{i'j'}$. For a comprehensive understanding of preemption, it is imperative to first define the neighboring node set.

We define neighboring nodes at a time step.
Let $\mathcal{N}_t(O_{ij})=\{ \mathcal{N}_{mt}(O_{ij}) \cup \mathcal{N}_{vt}(O_{ij}) \}$ be the neighboring nodes for $O_{ij}$ at time $t$, where $\mathcal{N}_{mt}(O_{ij})$ is the neighboring machines and $\mathcal{N}_{vt}(O_{ij})$ is the neighboring vehicles.
$\mathcal{N}_{mt}(O_{ij})$ denotes available machines of $O_{ij}$ at $t$.
Some machines among $\mathcal{M}_{ij}$ may be unavailable due to preemption from other operations, $\mathcal{N}_{mt}(O_{ij}) \subseteq \mathcal{M}_{ij}$.
$\mathcal{N}_{vt}(O_{ij})$ denotes current available vehicles, except for the transporting ones, $\mathcal{N}_{vt}(O_{ij}) \subseteq \mathcal{V}$.
Likewise, let $\mathcal{N}_t(M_{k})$ be neighboring operation nodes for machine $M_k$, and $\mathcal{N}_t(V_u)$ be neighboring operation nodes for vehicle $V_u$.

Neighboring nodes of an operation node depends on whether the neighboring nodes are working at time step $t$. 
For instance, consider the scenario where $\mathcal{N}_{mt}(O_{ij}) = \{ M_1, M_2, M_3 \}$, and the agent assigns $O_{i'j'}$ to $M_1$ at time $t$, without considering vehicle allocation for simplification. 
Due to the preemption of $M_1$ by another operation $O_{i'j'}$, the set of neighboring nodes changes to $\mathcal{N}_{m(t+1)}(O_{ij}) = \{M_2, M_3 \}$.
The same is true for the neighbors of machine and vehicle nodes.
A selection of initial neighboring nodes is sampled randomly, which is described in Sec.~\ref{sec:eval_instances}.

\subsection{Markov decision process}
We establish a Markov decision process (MDP) model for FJSPT.
At every time step $t$, the agent perceives the system state $s_t$ and selects an action $a_t$.
The action $a_t$ enables an unassigned operation to be processed on a free machine by conveying it with an available vehicle.
Subsequent to executing the action, the environment transitions to the next state $s_{t+1}$ and acquires reward $r_{t+1}$. This procedure continues until all operations are scheduled. The MDP model is explicitly detailed as follows.

\subsubsection{State} \label{sec:mdp_state}
At decision step $t$, state $s_t$ is a heterogeneous graph $\mathcal{H}_t(\mathcal{O} \cup \mathcal{M} \cup \mathcal{V}, \mathcal{C}, \mathcal{E}_{mt} \cup \mathcal{E}_{vt}^\text{off} \cup \mathcal{E}_{v}^\text{on})$.
In this graph, we define the raw features of nodes and edges.
Raw feature vector $\mu_{ij} \in \mathbb{R}^7$ of operation node $O_{ij}$ comprises 7 elements; 
\begin{itemize}
    \item Status: a binary value indicates 1 if $O_{ij}$ is scheduled until time $t$, otherwise 0.
    \item Number of neighboring machines: $|\mathcal{N}_{mt}(O_{ij})|$
    \item Number of neighboring vehicles: $|\mathcal{N}_{vt}(O_{ij})|$
    \item Processing time: $T_{ijk}^p$ if $O_{ij}$ is scheduled on $M_k$, otherwise, average processing time $\Bar{T}_{ij}^p$ for compatible machines $M_k \in \mathcal{M}_{ij}$, where $\Bar{T}_{ij}^p=\sum_{M_k\in \mathcal{M}_{ij}} T_{ijk}^p / |\mathcal{M}_{ij}|$.
    \item Number of unscheduled operations in job $J_i$: $n_i - |F_t(J_i)|$, where $F_t(J_i)$ is a set of finished operations in $J_i$ until time $t$.
    \item Job completion time: $C_i$. It has the actual value if $J_i$ is completed until $t$, otherwise, it indicates an estimation of the completion time, $\hat{C}_i = C_{ij'} + \sum_{O_{ij} \in J_i \setminus F_t(J_i)} \Bar{T}_{ij}^p$, where $O_{ij'}$ is the last finished operation until $t$.
    \item Start time: if $O_{ij}$ is scheduled until $t$, it indicates the actual start time of operation $O_{ij}$. Otherwise, it indicates the estimated start time of $O_{ij}$, $T_{ij}^s=C_{ij'} + \sum_{z=j'+1}^{j-1} \bar{T}^p_{iz}$, where $j'< j-1$, and $T_{ij}^s=C_{ij'}$, where $j'=j-1$.
\end{itemize}

Raw feature vector $\mu_k \in \mathbb{R}^4$ of machine $M_k$ consists of 4 elements; 
\begin{itemize}
    \item Status: a binary value indicates 1 if $M_k$ is processing at time $t$, otherwise 0.
    \item Number of neighboring operations: $|\mathcal{N}_{t}(M_k)|$.
    \item Available time: the time when $M_k$ completes all the allocated operations and can process new operations.
    \item Utilization: a ratio of $M_k$ usage time to the current time $t$.
\end{itemize}

Raw feature vector $\mu_{v} \in \mathbb{R}^4$ of vehicle $V_u$ consists of 4 elements; 
\begin{itemize}
    \item Status: a binary value indicates 1 if $V_u$ is transporting at time $t$, otherwise 0.
    \item Number of neighboring operations: $|\mathcal{N}_{t} (V_u)|$.
    \item Available time: the time when $V_u$ completes the allocated transportation operations and can transport new operations.
    \item Current location: $\mathcal{L}_t(V_u)$. We define candidate locations of the vehicle as a machine set, $\mathcal{L}_t(V_u)\in \mathcal{M}$.
\end{itemize}

Raw feature vector $\nu_{ijk}^m \in \mathbb{R}$ of compatible machine arc $E_{ijk}^m \in \mathcal{E}_m$ contains a single element: processing time $T_{ijk}^{p}$ for $O_{ij}$-$M_k$ pair.
Feature vector $\nu_{iju}^v \in \mathbb{R}$ of off-load vehicle arc $E_{iju}^v \in \mathcal{E}_v^\text{off}$ represents off-load transportation time $T_{iju}^t$, and vector $\nu_{kk'}^v \in \mathbb{R}$ of on-load vehicle arc $E_{kk'}^v \in \mathcal{E}_v^\text{on}$ represents on-load transportation time $T_{kk'}^t$.
We refer to the raw feature vectors of nodes and edges from paper~\cite{b10}.

\subsubsection{Action}
To address FJSPT, we define a composite action, $a_t=(O_{ij}, M_k, V_u)$, which constitutes operation selection, machine assignment and vehicle utilization. 
This implies that the operation $O_{ij}$ is assigned to the machine $M_k$, with $V_u$ transporting it. 
Specifically, action $a_t\in A_t$ is to select a feasible operation-machine-vehicle pair.
Feasible operation $O_{ij}$ implies its immediate predecessor $O_{i(j-1)}$ has been completed.
A feasible machine is an idle one among the compatible machines, $M_k \in \mathcal{M}_{ij}$.
A feasible vehicle is an idle one among all vehicles at time $t$, $V_u \in \mathcal{V}_t$.

\subsubsection{State transition}
Upon taking the action, the environment deterministically transitions to the next state $s_{t+1}$.
We first define the feasible next decision step $t+1$.
This step is determined by operation events, that is, the earliest release time of the new operation among the remaining feasible operations after time $t$.
At step $t+1$, the graph structure and node features are altered by the action $a_t=(O_{ij}, M_k, V_u)$, in ways such as node $O_{ij}$ retaining only one O-M-V pair in the graph, while other compatible pairs are removed, and specific features of nodes change as described in Section~\ref{sec:mdp_state}.

\subsubsection{Reward}
The objective is to learn how to schedule operations in such a way that the makespan is minimized.
We construct the reward function as the difference between the makespan corresponding to $s_t$ and $s_{t+1}$, $r(s_t, a_t, s_{t+1}) = C_{max}(s_{t}) - C_{max}(s_{t+1})$. 
In this context, we define $C_{LB}(O, s_t)$ as the lower bound of the estimated completion time of operation $O$ at state $s_t$. 
We compute this lower bound recursively, $C_{LB}(O_{ij}, s_t)=C_{LB}(O_{i(j-1)}, s_t) + \bar{T}_{ij}^p$, where $O_{ij}$ is an unscheduled operation in job $J_i$. 
If $O_{i(j-1)}$ is scheduled in $s_t$, $C_{LB}(O_{i(j-1)}, s_t)$ is updated to the actual completion time $C_{i(j-1)}$. 
We define the makespan in $s_t$ as the maximum lower bound of unscheduled operations for all jobs, $C_{max}(s_t) = \max_{i,j} \{ C_{LB}(O_{ij}, s_t) \}$. The makespan in the terminal state $s_{|O|}$ corresponds to the actual process makespan, $C_{max}(s_{|O|})=C_{max}$, since all operations are scheduled. 
When the discount factor $\gamma=1$, the cumulative reward is $G = \sum_{t=0}^{|O|}r(s_t, a_t, s_{t+1})=C_{max}(s_0) - C_{max}$, where $C_{max}(s_0)$ is a constant for a specific instance. 
Therefore, maximizing $G$ is equivalent to minimizing the makespan.

\subsection{Structure-aware heterogeneous encoder}
A key concept of the proposed encoder is to construct three sub-encoders (for operation, machine and vehicle) that separately aggregate the neighboring messages while considering node class.
The graph's structural similarity affects the scale generalization~\cite{b23}.
Sub-graph-based encoding methods contribute to improved scale generalization, as the sub-graph has higher structural similarity than that of the entire graph in large-scale graphs.
Consequently, we design the sub-graphs corresponding to node classs, and then the sub-encoder extracts the embedding features of the sub-graph.
Intuitively, from the perspective of machine nodes, the priority is selecting operation nodes with low processing time, while disregarding vehicle transportation time.
Conversely, from the perspective of vehicle nodes, the priority is selecting operation nodes with low transportation time, while disregarding machine processing time.
From the perspective of operation nodes, it should be assigned to the nodes with low transportation and processing time at the same time.
Following the local encoding of graph nodes, the global encoder integrates the messages from all nodes.

\subsubsection{Sub-encoders}
We develop the sub-encoders $\mathcal{F}_{O}, \mathcal{F}_{M}$ and $\mathcal{F}_{V}$ for operation, machine and vehicle node, respectively.
In contrast to traditional AM, which represents single-class nodes, the sub-encoder captures node embedding under different node classes and outputs both node and edge embeddings.
Let $h_{ij}^{(l)}$, $h_k^{(l)}$ and $h_u^{(l)}$ be the operation machine and vehicle node embedding vector, respectively, through layer $l\in \{ 1, ..., L-1 \}$. There are multiple $L-1$ attention layers.
The process of locally extracting relationship knowledge is that sub-encoder $\mathcal{F}_{X}^{(l)}$ generates updated node embedding vector $h_x^{(l)}$ of node $x \in X$ and its edge embedding vector $h_{xy}^{(l)}$ by aggregating knowledge of the self node embedding $h_x^{(l-1)} \in \mathbb{R}^{d_h}$ at the previous layer $l-1$, its neighboring node embedding $h_y^{(l-1)} \in \mathbb{R}^{d_h}$ with different node-class $y\in Y$, and their relationship (edge) $h_{xy}^{(l-1)} \in \mathbb{R}^{d_e}$.
Through this process, updated embedding vector $h_x^{(l)}$ from $\mathcal{F}_X^{(l)}$ includes local knowledge of its neighboring nodes and their relationship, and reflects more information from more relevant neighbors.
This is formulated as follows:
\begin{equation}
\begin{aligned}
    h_{ij}^{(l)}, & \{ h_{ijk}^{(l)}, h_{iju}^{(l)} | M_k \in \mathcal{N}_{mt}(O_{ij}), V_u \in \mathcal{N}_{vt}(O_{ij}) \} \\
    & = \mathcal{F}_{O}^{(l)}(h_{ij}^{(l-1)}, \{ h_k^{(l-1)}, h_{ijk}^{(l-1)} | M_k \in \mathcal{N}_{mt}(O_{ij}) \} \\
    & \cup \{ h_u^{(l-1)}, h_{iju}^{(l-1)} | V_u \in \mathcal{N}_{vt}(O_{ij})  \} ) \\
    h_{k}^{(l)}, & \{ h_{kij}^{(l)}, h_{kk'}^{(l)} | O_{ij} \in \mathcal{N}_t(M_k), M_{k'} \in \mathcal{M}_t\}  \\
    & = \mathcal{F}_{M}^{(l)}(h_{k}^{(l-1)}, \{ h_{ij}^{(l-1)}, h_{ijk}^{(l-1)} | O_{ij} \in \mathcal{N}_t(M_{k}) \} \\
    & \cup \{ h_{k'}^{(l-1)}, h_{kk'}^{(l-1)} | M_{k'} \in \mathcal{M}_t \}) \\
    h_u^{(l)}, & \{ h_{uij}^{(l)} | O_{ij} \in \mathcal{N}_t(V_u) \}  \\
    & = \mathcal{F}_{V}^{(l)}(h_{u}^{(l-1)}, \{ h_{ij}^{(l-1)}, h_{iju}^{(l-1)} | O_{ij} \in \mathcal{N}_t(V_u) \})
\end{aligned}
\end{equation}
where $h^{(l-1)}$ is a node embedding at layer $l-1$.
$h_{ijk}^{(l)}$ and $h_{kij}^{(l)}$ are embedding vectors for edge $O_{ij}$-$M_k$.
$h_{iju}^{(l)}$ and $h_{uij}^{(l)}$ are the vectors for edge $O_{ij}$-$V_u$.
$h_{kk'}^{(l)}$ is the vector between machines $M_k$ and $M_{k'}$.
The sub-encoder $\mathcal{F}_X^{(l)}$ is composed of heterogeneous multi-head attention (HMHA), add $\&$ normalization (AN), and feed-forward layer (FF). 
HMHA performs the knowledge aggregation process.

\begin{figure}[t]
    \centering
    \includegraphics[width=1.0\linewidth, , height=5cm]{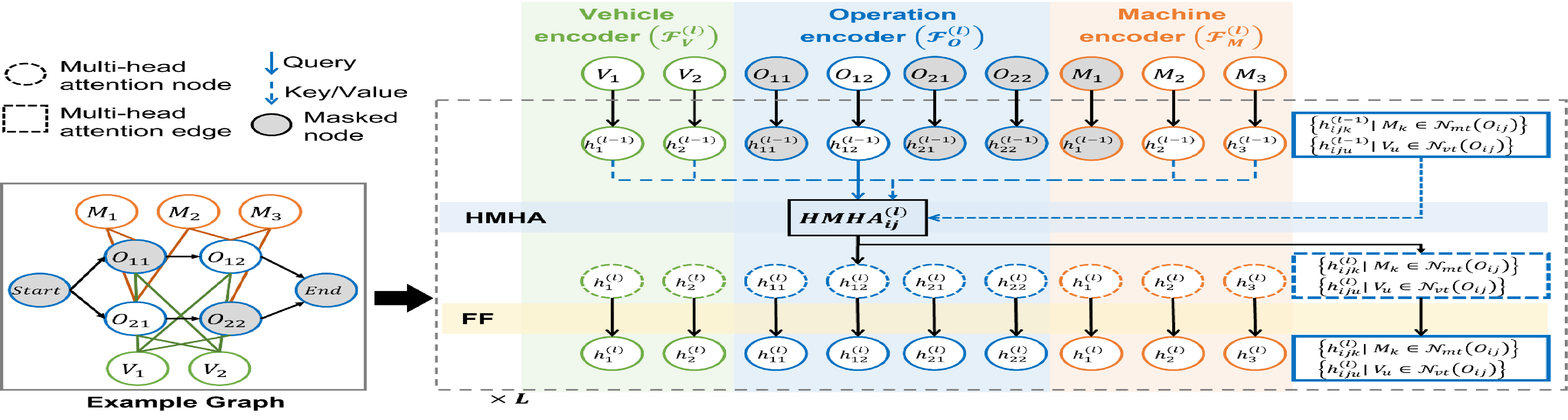}
    \caption{Heterogeneous encoder architecture. This figure shows the sub-encoding process of the given example graph. For simplicity, we only illustrate $\text{HMHA}_{ij}$ block between operation node $O_{ij}$ and its neighboring nodes in the heterogeneous multi-head attention layer. }
    \label{fig:encoder}
    \vspace{-10pt}
\end{figure}

We design HMHA block that embeds messages of different classes of nodes and their relationship (edge) based on AM~\cite{b8}.
The HMHA block learns attention scores between nodes of how much they relate to each other.
In FJSPT, an operation node and the corresponding compatible machine node with a low processing time may have a high attention score because this O-M pair contributes to the low makespan.
To this end, we incorporate the message-passing techniques between different node classes into the traditional AM, while taking into account edge attributes such as processing and transportation time.

For node embedding $h_x^{(l)}$, $\text{HMHA}_x$ block for node $x$ embeds messages of neighboring nodes $\{h_y^{(l-1)} | y\in \mathcal{N}(x) \subset Y \}$ and their edge messages $h_{xy}^{(l-1)}$.
With query $\text{q}_x$ for node $x$ and key $\text{k}_y$/value $\text{v}_y$ for neighboring node $y$ from equation~(\ref{eqn:qkv}), we computes the compatibility $\sigma_{xy}$ by following equation~(\ref{eqn:compatability}).
To include edge messages, we define an \textit{augmented compatibility} $\Tilde{\sigma}_{xy}$, which uses two-step linear transformation on the concatenation of compatibility $\sigma_{xy}$ and edge $\nu_{xy}$:
\begin{equation}
\begin{aligned}
    \Tilde{\sigma}_{xy} = W_{xy}^{e_2} \cdot \text{ReLU} \left( W_{xy}^{e_1} \left[ \sigma_{xy} \parallel h_{xy}^{(l-1)} \right] \right),
\end{aligned}
\end{equation}
where $W_{xy}^{e_1} \in \mathbb{R}^{d_z \times (1+d_e)}$ and $W_{xy}^{e_2} \in \mathbb{R}^{1 \times d_z}$ are trainable parameter matrix, $[\cdot \parallel \cdot]$ denotes a concatenation function, and ReLU is an activation function.
With the augmented compatibility, we calculate attention weights $\Bar{\sigma}_{xy}$ using equation~(\ref{eqn:score}), single-head node embedding $h_{x,z}'^{(l)}$ using equation~(\ref{eqn:single-head attention}), and finally updated node embedding $h_x^{(l)}$ using equation~(\ref{eqn:multi-head attention}).
Additionally, we compute the updated edge embedding $h_{xy}^{(l)}$ through a linear transformation, leveraging the compatibility $\Tilde{\sigma}_{xy}$:
\begin{equation}
\begin{aligned}
    h_{xy}^{(l)} = W_{xy}^{e_3} \Tilde{\sigma}_{xy},
\end{aligned}
\end{equation}
where $W_{xy}^{e_3}\in \mathbb{R}^{d_e \times 1}$ is the trainable parameter matrix.

Funtionally, the $\text{HMHA}_x^{(l)}$ block at layer $l$ takes neighboring node embeddings and edge embeddings as input, $\{ h_{y}^{(l-1)}, h_{xy}^{(l-1)} | y\in \mathcal{N}(x)\}$, and outputs the updated multi-head node embedding $h_x^{(l)}$ and their edge embeddings $h_{xy}^{(l)}$:
\begin{equation}
\begin{aligned}
    h_x^{(l)}, &\{ h_{xy}^{(l)} | y\in \mathcal{N}(x) \} \\
    & = \text{HMHA}_x^{(l)}(\{ h_{y}^{(l-1)}, h_{xy}^{(l-1)} | y\in \mathcal{N}(x)\}).
\end{aligned}
\end{equation}

To apply it in terms of operation node, the node embedding $h_{ij}^{(l)}$ aggregates neighboring node and edge messages from both machine and vehicle nodes:
\begin{equation}
\begin{aligned}
    h_{ij}^{(l)}, & \{ h_{ijk}^{(l)} | M_k \in \mathcal{N}_{mt}(O_{ij}) \} \cup \{h_{iju}^{(l)} | V_u \in \mathcal{N}_{vt}(O_{ij}) \} \\
    & = \text{HMHA}_{ij}^{(l)} ( \{ h_k^{(l-1)}, h_{ijk}^{(l-1)} | M_k \in \mathcal{N}_{mt}(O_{ij}) \} \\ 
    & \cup \{ h_u^{(l-1)}, h_{iju}^{(l-1)} | V_u \in \mathcal{N}_{vt}(O_{ij}) \}),
\end{aligned}
\end{equation}
where edge embedding $h_{ijk}$ captures processing time knowledge between $O_{ij}$ and $M_k$, and $h_{iju}$ captures off-load transportation time between $O_{ij}$ and $V_u$.

From the perspective of the machine node, the $\text{HMHA}_k^{(l)}$ block incorporates both messages of $h_{ijk}$ edge (processing time) and $h_{kk'}$ edge (on-load transportation time).
This can be expressed as follows:
\begin{equation}
\begin{aligned}
    h_k^{(l)}, & \{ h_{kij}^{(l)} | O_{ij} \in \mathcal{N}_t(M_k) \} \cup \{ h_{kk'}^{(l)} | M_{k'} \in \mathcal{M}_t\} \\
    & = \text{HMHA}_k^{(l)} ( \{ h_{ij}^{(l-1)}, h_{kij}^{(l-1)} | O_{ij} \in \mathcal{N}_t(M_k) \} \\
    & \cup \{ h_{k'}^{(l-1)}, h_{kk'}^{(l-1)} | M_{k'} \in \mathcal{M}_t \})
\end{aligned}
\end{equation}
where we denote $O_{ij}$-$M_k$ edge embedding as $h_{kij}$ to avoid confusion with $h_{ijk}$ in $\text{HMHA}_{ij}^{(l)}$ block, although they have the same values.
When the edge embedding is used for $\text{HMHA}_{ij}^{(l+1)}$ and $\text{HMHA}_{k}^{(l+1)}$ blocks at the next layer, we employ the sum of $h_{ijk}^{(l)}$ and $h_{kij}^{(l)}$ as the input.

Vehicle nodes have the relationship with operation nodes, the $\text{HMHA}_{u}^{(l)}$ block considers the off-load transportation time:
\begin{equation}
\begin{aligned}
    h_u^{(l)}, & \{ h_{uij}^{(l)} | O_{ij} \in \mathcal{N}_t(V_u) \} \\
    &= \text{HMHA}_u^{(l)} ( \{ h_{ij}^{(l-1)}, h_{uij}^{(l-1)} | O_{ij} \in \mathcal{N}_t(V_u) \})    
\end{aligned}
\end{equation}
where $h_{uij}^{(l)}$ is equivalent to $h_{iju}^{(l)}$ used in the $\text{HMHA}_{ij}^{(l)}$ block.

Feed-forward (FF) block in Fig.~\ref{fig:encoder} is implemented with two hidden layers:
\begin{equation}
\begin{aligned}
    \text{FF}(h^{(l)}_x) & = W^{\text{ff}_2}_{x} \cdot \text{ReLU}(W^{\text{ff}_1}_{x} h^{(l)}_x) \\
    \text{FF}(h^{(l)}_{xy}) & = W^{\text{ff}_2}_{xy} \cdot \text{ReLU}(W^{\text{ff}_1}_{xy} h^{(l)}_{xy}),
\end{aligned}
\end{equation}
where $W^{\text{ff}_1}_{x} \in \mathbb{R}^{d_{\text{ff}} \times d_h}$, $W^{\text{ff}_2}_{x} \in \mathbb{R}^{d_h \times d_{\text{ff}}}$, $W^{\text{ff}_1}_{xy} \in \mathbb{R}^{d_{\text{ff}} \times d_e}$ and $W^{\text{ff}_2}_{xy} \in \mathbb{R}^{d_e \times d_{\text{ff}}}$ are trainable matrix.
In addition, add $\&$ normalization (AN) block is implemented by using an instance normalization for stable and fast training~\cite{b6}.

\subsubsection{Global encoder}
After $L-1$ sub-encoding, the global encoder $\mathcal{F}_{G}^{(L)}$ incorporates messages of all nodes and edges.
Let $x$ be a graph node $x\in X= \mathcal{O} \cup \mathcal{M} \cup \mathcal{V}$.
The final node and edge embeddings from the heterogeneous encoder are obtained as follows:
\begin{equation}
\begin{aligned}
    h_x^{(L)}, & \{ h_{xy}^{(L)} | y \in \mathcal{N}_t(x) \} \\
    & = \mathcal{F}_{G}^{(L)}(h_{x}^{(L-1)}, \{ h_{y}^{(L-1)}, h_{xy}^{(L-1)} | y \in \mathcal{N}_t(x) \}),
\end{aligned}
\end{equation}
where $\mathcal{N}_t(x)$ is a neighboring node set of $x$ at time step $t$.
Likewise sub-encoders, $\mathcal{F}_{G}^{(L)}$ comprises equivalent $\text{HMHA}_x^{(L)}$, AN and FF blocks.
The $\text{HMHA}_x^{(L)}$ allows a node $x$ to incorporate messages of its neighboring nodes $y \in \mathcal{N}_t(x)$, as follows:
\begin{equation}
\begin{aligned}
    h_x^{(L)}, & \{ h_{xy}^{(L)} | y \in \mathcal{N}_t(x) \} \\
    &= \text{HMHA}_x^{(L)} ( \{ h_{y}^{(L-1)}, h_{xy}^{(L-1)} | y \in \mathcal{N}_t(x) \}).
\end{aligned}
\end{equation}

\subsection{Three-stage decoder}
Through sub-encoders with $L$ layers, we obtain node and edge embeddings; $ \{ h_{ij}^{(L)}, h_{k}^{(L)}, h_{u}^{(L)}, h_{ijk}^{(L)}, h_{iju}^{(L)}, h_{kk'}^{(L)} | O_{ij} \in \mathcal{O}, M_k \in \mathcal{M}, M_{k'} \in \mathcal{M},  V_u\in \mathcal{V} \}$.
With these embeddings, the proposed decoder determines an action $a_t = (O_{ij}, M_k, V_u$) of the operation-machine-vehicle pair at the decision step.
The decoder comprises three multi-head attention (MHA) sub-layers that generate the probability vectors for selecting operation, machine and vehicle nodes.
Each node element of action $a_t$ is sampled from the vectors.

\begin{figure}[t]
    \centering
    \includegraphics[width=0.9\linewidth, height=5.5cm]{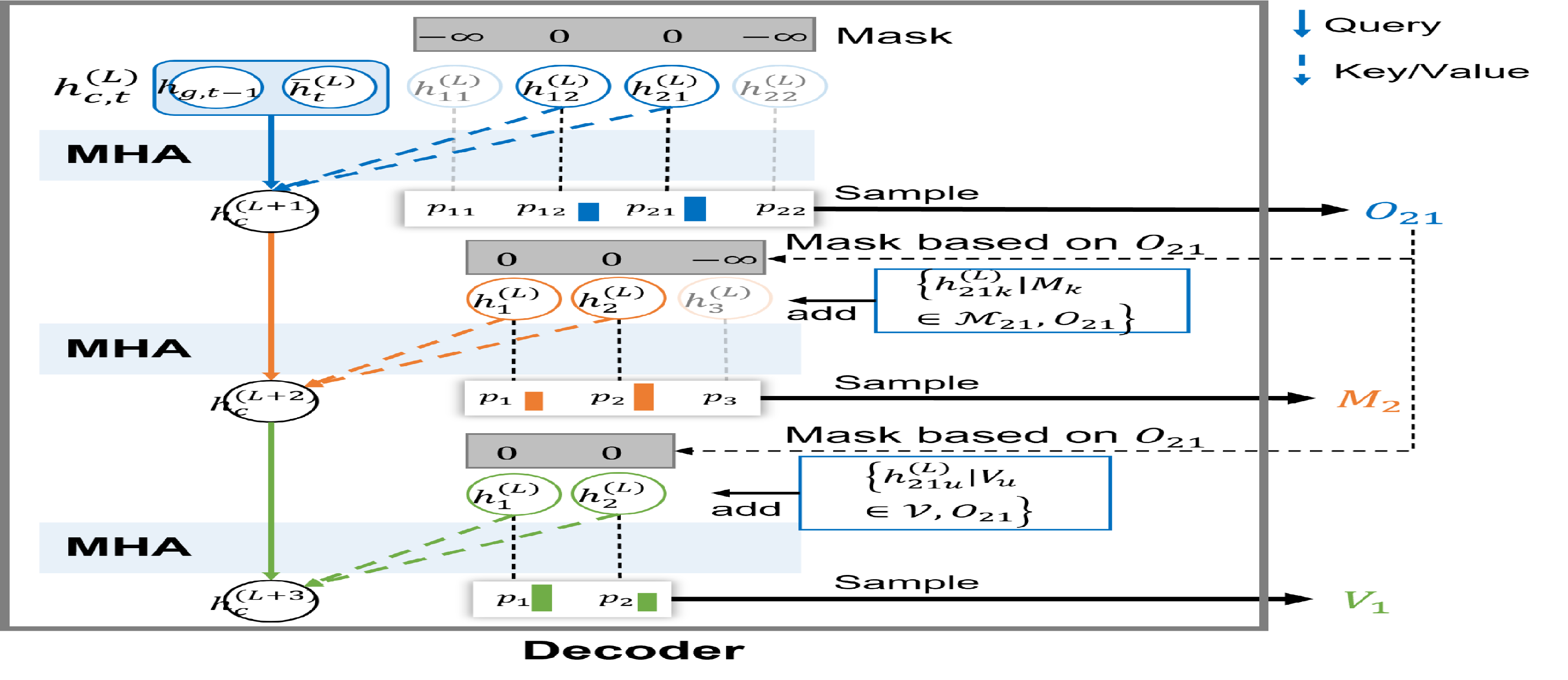}
    \caption{Three-stage decoder architecture.}
    \label{fig:decoder}
    \vspace{-10pt}
\end{figure}

\subsubsection{Operation node selection}
Initially, we construct a \textit{context} node $h_{c,t}^{(L)}$ to include the current graph embedding $\Bar{h}_t^{(L)}$ at step $t$ and last glimpse node embedding $h_{g, t-1}$ at step $t-1$:
\begin{equation}
\begin{aligned}
    h_{c,t}^{(L)} = \left[\Bar{h}_t^{(L)} \parallel h_{g,t-1} \right],
\end{aligned}
\end{equation}
where $h_{g, t-1}$ is related to the last selected nodes, which is defined in equation~(\ref{eqn:glimpse}).
We formulate the graph embedding as the mean of all node embeddings, $\Bar{h}^{(L)}=\frac{1}{|\mathcal{O}| + |\mathcal{M}| + |\mathcal{V}|} ( \sum_{i=1}^{n} \sum_{j=1}^{n_i} h_{ij}^{(L)} + \sum_{k=1}^{m} h_{k}^{(L)} + \sum_{u=1}^{v} h_{u}^{(L)})$.
For simple notation, we omit the subscript $t$, such as $h_{c}^{(L)}=h_{c,t}^{(L)}$.
Here, we need to determine which operation node is most related to the context node.
To this end, we use the basic multi-head attention computation of equation~(\ref{eqn:multi-head attention}) based on the context node to compute the node selection probability.
The context node $h_c^{(L+1)}$, which aggregates messages from operation embeddings, is expressed as follows:
\begin{equation}
\begin{aligned}
\label{eqn:mha_operation}
    h_c^{(L+1)} = \text{MHA}_c^{(L+1)}(\{ h_{ij}^{(L)} | O_{ij}\in \mathcal{O} \}),
\end{aligned}
\end{equation}
where the context node aggregates the knowledge of operation node embeddings $h_{ij}^{(L)}$ by utilizing query $\text{q}_c$ for $h_c^{(L)}$, and key $\text{k}_{ij}$/value $\text{v}_{ij}$ for $h_{ij}^{(L)}$, as derived from equation~(\ref{eqn:qkv}).

To compute the probability for operation selection, we add a single-head attention (SHA) layer ($Z=1$)~\cite{b8}.
In this layer, we compute the compatibility $\sigma_{cij}^{(L+1)}$ between context $h_c^{(L+1)}$ and operation embedding $h_{ij}^{(L)}$, likewise equation~(\ref{eqn:compatability}):
\begin{equation}
\begin{aligned}
\label{eqn:compatibility_SHA_oper}
    \sigma_{cij}^{(L+1)} = 
    \begin{cases}
        C \cdot \tanh \left( \frac{[h_c^{(L+1)}]^\mathrm{T} h_{ij}^{(L)}}{\sqrt{d_\text{k}}} \right)  & \text{if } O_{ij} \text{ is eligible} \\
        -\infty & \text{otherwise}
    \end{cases}
\end{aligned}
\end{equation}
where $C$ is set to 10 to clip the result for better exploration~\cite{b8}. 
Concurrently, to ensure feasibility, we dynamically mask non-eligible operations at each step with $-\infty$.
Completed operations and those out of sequence require masking.
Typically, these compatibilities are viewed as unnormalized log-probabilities (logits)~\cite{b8}.
Ultimately, by employing softmax, the probability distribution for operation selection is calculated as follows:
\begin{equation}
\begin{aligned}
    Pr(O_{ij}|s_t)=\frac{e^{\sigma_{cij}^{(L+1)}}}{\sum_{i'=1}^{n} \sum_{j'=1}^{n_{i'}} e^{\sigma_{ci'j'}^{(L+1)}}}.
\end{aligned}
\end{equation}
We can select an operation node $O_{ij}$ sampled from the distribution.

\subsubsection{Machine node selection}
Considering the selected operation node $O_{ij}$, we compute the machine selection probability distribution. 
To involve knowledge of the selected operation $O_{ij}$ to the machine selection, we utilize edge embedding $h_{ijk}^{(L)}$ by adding it to the MHA input:
\begin{equation}
\begin{aligned}
\label{eqn:mha_machine}
    h_c^{(L+2)} = \text{MHA}_c^{(L+2)}(\{ h_{k}^{(L)} + h_{ijk}^{(L)} \\
    | M_k\in \mathcal{M}_{ij}, \text{selected } O_{ij} \}),
\end{aligned}
\end{equation}
where the used query $\text{q}_c$ and key $\text{k}_k$/value $\text{v}_k$ correspond to context embedding $h_{c}^{(L+1)}$ and machine embedding $h_{k}^{(L)}$, respectively.
The log-probability for machine nodes is calculated by determining the compatibility $\sigma_{ck}^{(L+2)}$ between the context embedding $h_{c}^{(L+2)}$ and the machine embedding $h_{k}^{(L)}$ using equation~(\ref{eqn:compatibility_SHA_oper}).
The probability distribution of machine selection is expressed as follows:
\begin{equation}
\begin{aligned}
    Pr(M_k|s_t, O_{ij})=\frac{e^{\sigma_{ck}^{(L+2)}}}{\sum_{k'=1}^{m} e^{\sigma_{ck'}^{(L+2)}}}.
\end{aligned}
\end{equation}
We sample machine node $M_k$ from the distribution.

\subsubsection{Vehicle node selection}
Similar to the previous computation, we first compute the context embedding $h_c^{(L+3)}$ for vehicle nodes at the $\text{MHA}_c^{(L+3)}$ layer.
In the heterogeneous graph structure, vehicle nodes are adjacent to operation nodes, and thus we define the layer by adding edge embedding $h_{iju}^{(L)}$ for the selected $O_{ij}$ into its input:
\begin{equation}
\begin{aligned}
\label{eqn:mha_vehicle}
    h_c^{(L+3)} = \text{MHA}_c^{(L+3)}(\{ h_{u}^{(L)} + h_{iju}^{(L)} \\
    | V_u\in \mathcal{V}, \text{selected } O_{ij} \}),
\end{aligned}
\end{equation}
where the used query $\text{q}_c$ and key $\text{k}_u$/value $\text{v}_u$ correspond to context embedding $h_{c}^{(L+2)}$ and vehicle embedding $h_{u}^{(L)}$, respectively.
Finally, we define the probability distribution of vehicle selection as follows:
\begin{equation}
\begin{aligned}
    Pr(V_u|s_t, O_{ij})=\frac{e^{\sigma_{cu}^{(L+3)}}}{\sum_{u'=1}^{v} e^{\sigma_{cu'}^{(L+3)}}},
\end{aligned}
\end{equation}
where we compute compatibility $\sigma_{cu}^{(L+3)}$ by using $h_c^{(L+3)}$ and $h_{u}^{(L)}$ using equation~(\ref{eqn:compatibility_SHA_oper}).
After selecting O-M-V nodes, we update the glimpse node embedding at step $t$ by summing up the selected node embeddings:
\begin{equation}
\begin{aligned}
\label{eqn:glimpse}
    h_{g,t}= h_{ij}^{(L)} + h_k^{(L)} + h_{u}^{(L)}.
\end{aligned}
\end{equation}

As a result, the proposed \our{} module generates node and edge embeddings by encoding state $s_t$, and then generates the composite action $a_t$ by decoding the embeddings.
We can define the probability of the composite action:
\begin{equation}
\begin{aligned}
    Pr(a_t|s_t) = Pr(O_{ij}|s_t) Pr(M_k|s_t, O_{ij}) Pr(V_u|s_t, O_{ij}).
\end{aligned}
\end{equation}

\subsection{RL algorithm}
Algorithm~\ref{alg:rl_algorithm} describes the training process of the \our{} module.
We employ a policy-gradient method to update the policy $\pi_{\theta}(a_t|s_t)$~\cite{b1}.
In this study, the policy corresponds to encoder-decoder models, and we denote the policy parameter $\theta$ simply as all of the trainable parameter matrices utilized in the encoder-decoder models.
Within an episode, given policy $\pi_\theta$, we record the sequence of state, action, and reward samples, $\tau=(s_0, a_0, r_0, ..., s_T, a_T, r_T)$.
We can compute the total return $G(\tau)=\sum_{t=0}^{T}r_t$ of the trajectory.
Our objective is to maximize the objective $\mathit{J}(\theta) = \mathbb{E}_{\pi_\theta}[G(\tau)]$, which corresponds to minimizing the makespan.
We use REINFORCE algorithm~\cite{b2} to train the policy $\pi_\theta$, as it has been proven effective for attention-based end-to-end learning~\cite{b8, b11}.

\begin{algorithm}[t]
\caption{Reinforcement Learning Algorithm}\label{alg:rl_algorithm}
\begin{algorithmic}[1]
\STATE \textbf{Input:} number of epoches $E$, number of episodes per epoch $E'$, batch size $B$, policy network $\pi_\theta$

\FOR{$epoch=1, ..., E$}
    \IF{$epoch \mod 20 = 0$}
        \STATE Generate a new batch of $B$ FJSPT instances
    \ENDIF
    \STATE $epi = 0$
    \WHILE{$epi < E'$} 
        \FOR{$b=1, ..., B$}
            \STATE Initial state $s_0^b$ based on instance $b$ 
            \STATE $t=0$ 
            \WHILE{$s_t^b$ is not terminal}
                \STATE $a_t^b  \sim \pi_\theta(\cdot|s_t^b)$ // encoding and decoding
                \STATE Receive $r_t^b$ and transit to next state
                \STATE $t=t+1$
            \ENDWHILE
            \STATE $G(\tau^b)=\sum_t r_t^b$
            \STATE Receive baseline return $G_{base}^b$ using Greedy Rollout policy
        \ENDFOR
        \STATE $\nabla_\theta \mathit{J}(\theta) \leftarrow \frac{1}{B} \sum_{b=1}^{B} (G(\tau^b) - G_{base}^b) \nabla_\theta \log \pi_\theta(\tau^b)$
        \STATE Update $\theta$ using $\nabla_\theta \mathit{J}(\theta)$
        \STATE $epi = epi + B$
    \ENDWHILE
    
\ENDFOR

\end{algorithmic}
\end{algorithm}

We use REINFORCE algorithm~\cite{b2} to train the policy $\pi_\theta$, as it has been proven effective for attention-based end-to-end learning~\cite{b8, b11}.
The parameter $\theta$ is optimized as follows:
\begin{equation}
\begin{aligned}
    \nabla_\theta \mathit{J}(\theta) \leftarrow \frac{1}{B} \sum_{b=1}^{B} (G(\tau^b) - G_{base}^b) \nabla_\theta \log \pi_\theta(\tau^b),
\end{aligned}
\end{equation}
where $\pi_\theta(\tau^b)= \prod_{t=0}^{T} \pi_\theta(a_t^b|s_t^b)$ is the probability of the sequence actions during batch $b$ trajectory (episode), and $\tau^b$ is batch $b$ trajectory samples.
To reduce the variance for the gradient, we use the deterministic greedy rollout baseline $G_{base}^b$ for the batch $b$, which selects nodes with maximum probability on the distribution $Pr(O_{ij}|s_t)$, $Pr(M_k|s_t, O_{ij})$ and $Pr(V_u| s_t, O_{ij})$.
The detailed training process is described in Algorithm~\ref{alg:rl_algorithm}.
Every 20 epochs, we generate new batch $B$ instances with different graph topology for given instance parameters ($n \times m \times v$).
Each batch includes a different number of operations per job, number of compatible machines for each operation, and processing/transportation times, which will be described in Sec.~\ref{sec:eval_instances}.

\section{Performance evaluation}
In this section, we evaluate the effectiveness of our proposed method (\our{}) from two perspectives: makespan and scale generalization.

\subsection{Experimental settings}

\subsubsection{Evaluation instances} \label{sec:eval_instances}
Similar to most related studies~\cite{b4, b9, b10}, we generate synthetic FJSPT instances for training and testing.
To generate random instances under given instance parameters ($n, m \text{ and }v$), we sample an instance from the uniform distribution, where the number of operations for each job is sampled in proportion to the number of machines, $n_i \sim $ U($0.8|\mathcal{M}|, 1.2|\mathcal{M}|$), and the number of compatible machines for each operation is sampled from the distribution, $|\mathcal{M}_{ij}| \sim \text{U}(1, |\mathcal{M}|$).
Processing time $T_{ijk}^p$ for $O_{ij}$-$M_k$ pair and transportation time $T_{kk'}^t$ between $M_k$ and $M_{k'}$ are sampled from $\text{U}(0.8\Bar{T}_{ij}^p, 1.2\Bar{T}_{ij}^p)$ and $\text{U}(0.8\Bar{T}_{kk'}^t, 1.2\Bar{T}_{kk'}^t)$, respectively.
Here, the average processing time $\Bar{T}_{ij}^p$ and average transportation time $\Bar{T}_{kk'}^t$ are sampled from $\text{U}(1, 30)$ and $\text{U}(1, 20)$, respectively.
To assess the makespan optimization performance, we evaluate our model on four small-scale instances of $n\times m \times v$ (5$\times$3$\times$3, 10$\times$3$\times$6, 10$\times$6$\times$3, 10$\times$6$\times$6), where the number of operations in each instance is in the range $[10, 70]$.
When we test the performance of the methods, we generate 100 different instances for each size and calculate the average of the obtained results.
Furthermore, we conduct tests our method on various benchmark datasets, which will be described in detail in Section~\ref{sec:benchmark_test}.

\subsubsection{Configuration}
The \our{} model is constructed by stacking $L=2$ encoding layers. The embedding dimension of nodes and edges, $d_h$ and $d_e$, is set to 128 and 1, respectively.
Additionally, we use $d_z=16$ for calculating augmented compatibility and $d_{\text{ff}}=512$ in the "Feed-forward" blocks.
The (H)MHA blocks employed in both the encoder and decoder utilize $Z=8$ attention heads. Each attention head processes query, key, and value as 8-dimensional vectors, denoted as $d_{\text{q}} = d_{\text{k}} = d_{\text{v}} = 8$.
To optimize the model, we employ the Adam optimizer with a learning rate of $2\times 10^{-4}$ and use a batch size $B=50$.
In the training process, an epoch corresponds to the training of the model on $E'=1,000$ episodes. 
We train $E=1,000$ epochs for given instance parameters $n \times m \times v$.

\subsubsection{Baselines}
We compare the performance of the proposed method with several baseline algorithms:
\begin{itemize}
    \item Shortest processing time first (SPT): dispatching rule.
    \item Longest processing time first (LPT): dispatching rule.
    \item First in first out (FIFO): dispatching rule.
    \item MatNet~\cite{b11}: A DRL-based algorithm for solving FJSPT has not been developed entirely, and this aims only to solve FJSP. In light of this, we augment these algorithms with a simple vehicle selection mechanism, namely the nearest vehicle selection (NVS) method, to make them applicable to FJSPT.
    \item Heterogeneous graph neural network (HGNN)~\cite{b10}: DRL-based algorithm. We augment the NVS method to HGNN.
    \item Improved genetic algorithm (IGA)~\cite{b24}: meta-heuristic algorithm. This method solves FJSPT based on a genetic algorithm.
\end{itemize}

\subsection{Makespan optimization}
\begin{figure}[t]
    \centering
    \includegraphics[width=0.6\linewidth]{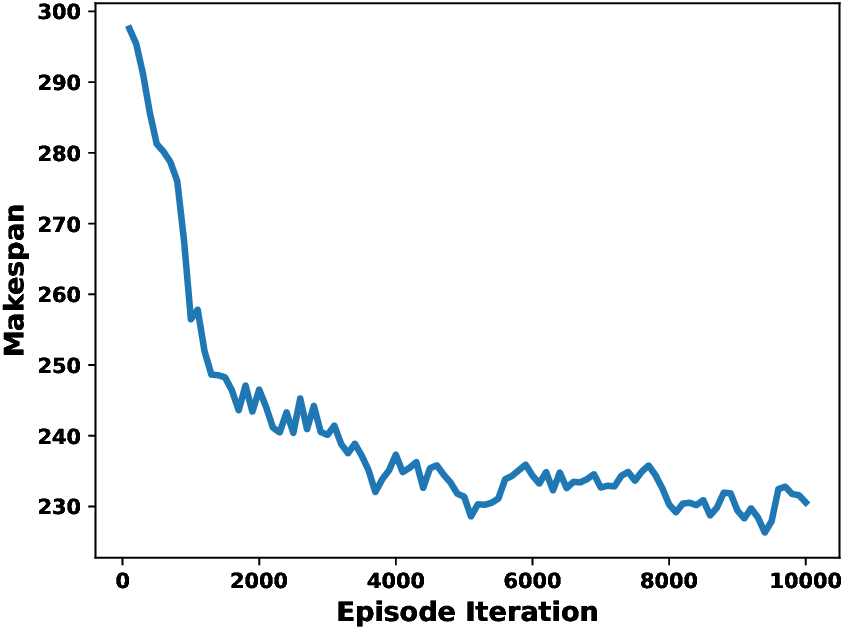}
    \caption{Training curve on 10$\times$6$\times$6 instances.}
    \label{fig:training_curve}
\end{figure}

To verify the effectiveness of the proposed method in finding near-optimal solutions, we evaluate the performance in terms of makespan $C_{max}$.
The training process of the proposed method is stable and converges for all instances. To illustrate this, in Fig.~\ref{fig:training_curve}, we plot the average makespan for 10 validation runs every 100 episode iterations on a 10×6×6 instance.
It is evident that the DRL agent is proficient in acquiring a high-quality scheduling policy from scratch, leveraging its own problem-solving experiences.

Next, after undergoing sufficient training consisting of 1,000 epochs, we evaluate its performance on synthetic instances using the trained model, making comparisons with baseline algorithms.
Table~\ref{tab:makespan_smallscale} shows the results of all methods against four instances, where Gap denotes the relative gap percentage between the method's makespan $C_{max}$ and makespan $C_{max}^{BS}$ of the best solution (not necessarily optimal) for the instance:
\begin{equation}
\begin{aligned}
    Gap = \left( \frac{C_{max}}{C_{max}^{BS}}-1 \right) \times 100\%.
\end{aligned}
\end{equation}
The bold character in the table denotes the best result for each instance.

For instances 5$\times$3$\times$3 and 10$\times$6$\times$6, \our{} provides the best solutions. Especially, it can obtain up to 24\% gap and 29\% gap when compared with other DRL-based methods and dispatch rules, respectively. 
For 10$\times$3$\times$6 instance, the meta-heuristic method (IGA) yields the best solution. 
However, it is notable that this method necessitates extensive computational resources.
Our proposed method, in contrast, obtains a near-best solution (with a 1\% gap) while demanding a more reasonable computational cost.
In the case of 10$\times$6$\times$3 instance, which is characterized by an insufficient number of vehicles, other DRL-based methods such as MatNet and HGNN achieve the best solution while requiring relatively lower computation times
However, these methods experience degraded performance in other instances.
Consequently, the proposed method consistently outperforms dispatch rules, meta-heuristic and existing DRL-based methods in small-scale instances.

\begin{table*}[t]
    \centering
    \footnotesize
    \begin{tabular}{ccccccccccccccc}
        \Xhline{3\arrayrulewidth}
        \multirow{4}{*}{\shortstack{Graph\\ Size}} 
        & \multicolumn{14}{c}{Methods} \\
        & \multicolumn{2}{c}{SPT} & \multicolumn{2}{c}{LPT} & \multicolumn{2}{c}{FIFO} & \multicolumn{2}{c}{IGA} & 
        \multicolumn{2}{c}{HGNN} & \multicolumn{2}{c}{MatNet} & \multicolumn{2}{c}{\our{} (Ours)} \\
        \cmidrule(lr){2-3} \cmidrule(lr){4-5} \cmidrule(lr){6-7} \cmidrule(lr){8-9} \cmidrule(lr){10-11} \cmidrule(lr){12-13} \cmidrule(lr){14-15}
        & \multirow{2}{*}{\shortstack{$C_{max}$ \\ (Time)}} & \multirow{2}{*}{Gap} & \multirow{2}{*}{\shortstack{$C_{max}$ \\ (Time)}} & \multirow{2}{*}{Gap} & \multirow{2}{*}{\shortstack{$C_{max}$ \\ (Time)}} & \multirow{2}{*}{Gap} & \multirow{2}{*}{\shortstack{$C_{max}$ \\ (Time)}} & \multirow{2}{*}{Gap} & \multirow{2}{*}{\shortstack{$C_{max}$ \\ (Time)}} & \multirow{2}{*}{Gap} & \multirow{2}{*}{\shortstack{$C_{max}$ \\ (Time)}} & \multirow{2}{*}{Gap} & \multirow{2}{*}{\shortstack{$C_{max}$ \\ (Time)}} & \multirow{2}{*}{Gap} \\
        \\
        \hline
        \multirow{2}{*}{5$\times$3$\times$3}
        & 102.75 & \multirow{2}{*}{12\%} & 107.95 & \multirow{2}{*}{26\%} & 105.3 & \multirow{2}{*}{23\%} & 109.95 & \multirow{2}{*}{28\%} & 105.95 & \multirow{2}{*}{24\%} & 96.55 & \multirow{2}{*}{13\%} & \textbf{85.65} & \multirow{2}{*}{\textbf{0\%}} \\
        & \textbf{(0.12s)} & & (0.12s) & & (0.13s) & & (272.51s) & & (0.13s) & & (0.34s) & & (0.2s) \\
        \hline

        \multirow{2}{*}{10$\times$3$\times$6}
        & 182.75 & \multirow{2}{*}{20\%} & 183.75 & \multirow{2}{*}{21\%} & 189.7 & \multirow{2}{*}{25\%} & 175.0 & \multirow{2}{*}{15\%} & 183.55 & \multirow{2}{*}{21\%} & 170.55 & \multirow{2}{*}{12\%} & \textbf{152.05} & \multirow{2}{*}{\textbf{0\%}}  \\
        & \textbf{(0.22s)} & & (0.22s) & & (0.61s) & & (569.8s) & & (0.25s) & & (0.32s) & & (0.4s) \\
        \hline

        \multirow{2}{*}{10$\times$6$\times$3}
        & 422.15 & \multirow{2}{*}{37\%} & 446.75 & \multirow{2}{*}{45\%} & 420.2 & \multirow{2}{*}{37\%} & 332.9 & \multirow{2}{*}{8\%} & \textbf{307.45} & \multirow{2}{*}{\textbf{0\%}} & 371.65 & \multirow{2}{*}{21\%} & 318.7 & \multirow{2}{*}{4\%} \\
        & (1.03s) & & (1.01s) & & (1.07s) & & (1005.5s) & & \textbf{(0.89s)} & & (1.62s) & & (1.33s) \\
        \hline

        \multirow{2}{*}{10$\times$6$\times$6}
        & 283.7 & \multirow{2}{*}{22\%} & 301.7 & \multirow{2}{*}{29\%} & 275.3 & \multirow{2}{*}{18\%} & 260.9 & \multirow{2}{*}{12\%} & 274.9 & \multirow{2}{*}{18\%} & 234.65 & \multirow{2}{*}{1\%} & \textbf{233.5} & \multirow{2}{*}{\textbf{0\%}} \\
        & (0.59s) & & \textbf{(0.58s)} & & (0.6s) & & (1005.91s) & & (0.62s) & & (1.02s) & & (0.95s) \\
        
        \Xhline{3\arrayrulewidth}
    \end{tabular}
    \caption{Makespance performances on small-scale instances. }
    \label{tab:makespan_smallscale}
    \vspace{-15pt}
\end{table*}

\subsection{Scale generalization}
Changes in the manufacturing environment on real shop floor, such as new jobs, machines and AGVs insertion, or machines/AGVs breakdown, are regarded as instance scale changes~\cite{b10, b37}.
To validate how well the proposed model generates schedules in the changing environment (scale generalization capability), we train the model on the small-scale instance 10$\times$6$\times$6 and subsequently test it on unseen-before large-scale instances (20$\times$10$\times$10, 30$\times$15$\times$15, 40$\times$20$\times$20, 50$\times$25$\times$25) where the number of operations in instances is in the range $[160, 1000]$.

Table~\ref{tab:makespan_largescale} shows that the proposed method significantly outperforms dispatch rules, meta-heuristic and existing DRL-based approaches, obtaining the best solutions (with a 0\% gap) across all instances.
It's noteworthy that the gap performance of the \our{} model amplifies as the size of the graph increases. 
Specifically, it achieves an increased gap ranging [24\%, 28\%] (from 24\% in instance 20$\times$10$\times$10 to 28\% in instance 50$\times$25$\times$25) for SPT, [30\%, 36\%] for LPT, [26\%, 31\%] for FIFO, [41\%, 54\%] for IGA and [21\%, 25\%] for HGNN.
Consequently, the proposed method proves capable of finding the best solutions in a variety of unseen-before instances.
This capability makes it highly valuable in the dynamic manufacturing environment subject to changes such as the addition or breakdown of machines/vehicles or the insertion of new jobs, as these changes directly correlate with changes in the graph scale.

\begin{table*}[t]
    \centering
    \footnotesize
    \scalebox{1.0}{
    \begin{tabular}{ccccccccccccccc}
        \Xhline{3\arrayrulewidth}
        \multirow{4}{*}{\shortstack{Graph\\ Size}} 
        & \multicolumn{14}{c}{Methods} \\
        & \multicolumn{2}{c}{SPT} & \multicolumn{2}{c}{LPT} & \multicolumn{2}{c}{FIFO} & \multicolumn{2}{c}{IGA} & 
        \multicolumn{2}{c}{HGNN} & \multicolumn{2}{c}{MatNet} & \multicolumn{2}{c}{\our{} (Ours)} \\
        \cmidrule(lr){2-3} \cmidrule(lr){4-5} \cmidrule(lr){6-7} \cmidrule(lr){8-9} \cmidrule(lr){10-11} \cmidrule(lr){12-13} \cmidrule(lr){14-15}
        & \multirow{2}{*}{\shortstack{$C_{max}$ \\ (Time)}} & \multirow{2}{*}{Gap} & \multirow{2}{*}{\shortstack{$C_{max}$ \\ (Time)}} & \multirow{2}{*}{Gap} & \multirow{2}{*}{\shortstack{$C_{max}$ \\ (Time)}} & \multirow{2}{*}{Gap} & \multirow{2}{*}{\shortstack{$C_{max}$ \\ (Time)}} & \multirow{2}{*}{Gap} & \multirow{2}{*}{\shortstack{$C_{max}$ \\ (Time)}} & \multirow{2}{*}{Gap} & \multirow{2}{*}{\shortstack{$C_{max}$ \\ (Time)}} & \multirow{2}{*}{Gap} & \multirow{2}{*}{\shortstack{$C_{max}$ \\ (Time)}} & \multirow{2}{*}{Gap} \\
        \\

        \hline
        \multirow{2}{*}{20$\times$10$\times$10}
        & 632.6 & \multirow{2}{*}{24\%} & 669.85 & \multirow{2}{*}{31\%} & 642.15 & \multirow{2}{*}{25\%} & 692.75 & \multirow{2}{*}{35\%} & 635.85 & \multirow{2}{*}{24\%} & 555.05 & \multirow{2}{*}{8\%} & \textbf{511.7} & \multirow{2}{*}{\textbf{0\%}} \\
        & \textbf{(1.83s)} & & (1.83s) & & (1.97s) & & (1020.75s) & & (2.06s) & & (2.62s) & & (3.4s) \\
        \hline

        \multirow{2}{*}{30$\times$15$\times$15}
        & 937.6 & \multirow{2}{*}{26\%} & 977.4 & \multirow{2}{*}{32\%} & 926.35 & \multirow{2}{*}{25\%} & 1108.45 & \multirow{2}{*}{49\%} & 904.65 & \multirow{2}{*}{22\%} & 800.05 & \multirow{2}{*}{8\%} & \textbf{742.0} & \multirow{2}{*}{\textbf{0\%}}  \\
        & \textbf{(3.97s)} & & (3.97s) & & (4.25s) & & (1052.05s) & & (6.35s) & & (5.77s) & & (8.04s) \\
        \hline

        \multirow{2}{*}{40$\times$20$\times$20}
        & 1206.05 & \multirow{2}{*}{25\%} & 1287.15 & \multirow{2}{*}{33\%} & 1219.05 & \multirow{2}{*}{26\%} & 1497.3 & \multirow{2}{*}{55\%} & 1181.9 & \multirow{2}{*}{22\%} & 1028.62 & \multirow{2}{*}{6\%} & \textbf{968.15} & \multirow{2}{*}{\textbf{0\%}} \\
        & (13.29s) & & \textbf{(13.27s)} & & (13.79s) & & (1101.04s) & & (45.16s) & & (20.17s) & & (32.27s) \\
        \hline

        \multirow{2}{*}{50$\times$25$\times$25}
        & 1572.35 & \multirow{2}{*}{26\%} & 1660.05 & \multirow{2}{*}{33\%} & 1580.5 & \multirow{2}{*}{26\%} & 2009.1 & \multirow{2}{*}{61\%} & 1510.4 & \multirow{2}{*}{21\%} & 1337.5 & \multirow{2}{*}{7\%} & \textbf{1251.4} & \multirow{2}{*}{\textbf{0\%}} \\
        & (11.69s) & & \textbf{(11.6s)} & & (12.38s) & & (1186.63s) & & (65.3s) & & (16.74s) & & (33.06s) \\
        
        \Xhline{3\arrayrulewidth}
    \end{tabular}
    }
    \caption{Scale generalization performances on large-scale instances.}
    \label{tab:makespan_largescale}
    \vspace{-15pt}
\end{table*}

\begin{table*}[t]
    \centering
    \footnotesize
    \scalebox{1.0}{
    
    \begin{tabular}{ccccccccccccccc}
        \Xhline{3\arrayrulewidth}
        \multirow{4}{*}{\shortstack{Graph\\ Size}} 
        & \multicolumn{14}{c}{Methods} \\
        & \multicolumn{2}{c}{SPT} & \multicolumn{2}{c}{LPT} & \multicolumn{2}{c}{FIFO} & \multicolumn{2}{c}{IGA} & 
        \multicolumn{2}{c}{HGNN} & \multicolumn{2}{c}{MatNet} & \multicolumn{2}{c}{\our{} (Ours)} \\
        \cmidrule(lr){2-3} \cmidrule(lr){4-5} \cmidrule(lr){6-7} \cmidrule(lr){8-9} \cmidrule(lr){10-11} \cmidrule(lr){12-13} \cmidrule(lr){14-15}
        & \multirow{2}{*}{\shortstack{$C_{max}$ \\ (Time)}} & \multirow{2}{*}{Gap} & \multirow{2}{*}{\shortstack{$C_{max}$ \\ (Time)}} & \multirow{2}{*}{Gap} & \multirow{2}{*}{\shortstack{$C_{max}$ \\ (Time)}} & \multirow{2}{*}{Gap} & \multirow{2}{*}{\shortstack{$C_{max}$ \\ (Time)}} & \multirow{2}{*}{Gap} & \multirow{2}{*}{\shortstack{$C_{max}$ \\ (Time)}} & \multirow{2}{*}{Gap} & \multirow{2}{*}{\shortstack{$C_{max}$ \\ (Time)}} & \multirow{2}{*}{Gap} & \multirow{2}{*}{\shortstack{$C_{max}$ \\ (Time)}} & \multirow{2}{*}{Gap} \\
        \\
        \hline
        \multirow{2}{*}{MKT01}
        & 175 & \multirow{2}{*}{34\%} & 198 & \multirow{2}{*}{51\%} & 146 & \multirow{2}{*}{11\%} & 141 & \multirow{2}{*}{8\%} & \textbf{131} & \multirow{2}{*}{\textbf{0\%}} & 159 & \multirow{2}{*}{21\%} & 153 & \multirow{2}{*}{17\%} \\
        & \textbf{(0.2s)} & & (0.2s) & & (0.24s) & & (1008.31s) & & (0.34s) & & (0.58s) & & (0.64s) \\
        \hline

        \multirow{2}{*}{MKT02}
        & 140 & \multirow{2}{*}{35\%} & 141 & \multirow{2}{*}{36\%} & 143 & \multirow{2}{*}{38\%} & 105 & \multirow{2}{*}{1\%} & 123 & \multirow{2}{*}{18\%} & 110 & \multirow{2}{*}{6\%} & \textbf{104} & \multirow{2}{*}{\textbf{0\%}} \\
        & \textbf{(0.2s)} & & (0.2s) & & (0.24s) & & (1009.24s) & & (0.34s) & & (0.39s) & & (0.64s) \\
        \hline

        \multirow{2}{*}{MKT03}
        & 415 & \multirow{2}{*}{55\%} & 453 & \multirow{2}{*}{70\%} & 331 & \multirow{2}{*}{24\%} & 335 & \multirow{2}{*}{25\%} & 306 & \multirow{2}{*}{15\%} & 314 & \multirow{2}{*}{18\%} & \textbf{267} & \multirow{2}{*}{\textbf{0\%}} \\
        & \textbf{(0.51s)} & & (0.52s) & & (0.61s) & & (1037.84s) & & (0.96s) & & (1.02s) & & (1.67s) \\
        \hline

        \multirow{2}{*}{MKT04}
        & 155 & \multirow{2}{*}{12\%} & 191 & \multirow{2}{*}{37\%} & 173 & \multirow{2}{*}{24\%} & 144 & \multirow{2}{*}{4\%} & 160 & \multirow{2}{*}{15\%} & 147 & \multirow{2}{*}{6\%} & \textbf{139} & \multirow{2}{*}{\textbf{0\%}} \\
        & \textbf{(0.3s)} & & (0.31s) & & (0.36s) & & (1011.07s) & & (0.52s) & & (0.61s) & & (1.0s) \\
        \hline

        \multirow{2}{*}{MKT05}
        & 473 & \multirow{2}{*}{26\%} & 442 & \multirow{2}{*}{18\%} & 461 & \multirow{2}{*}{23\%} & 423 & \multirow{2}{*}{13\%} & 388 & \multirow{2}{*}{4\%} & 412 & \multirow{2}{*}{10\%} & \textbf{374} & \multirow{2}{*}{\textbf{0\%}} \\
        & \textbf{(0.37s)} & & (0.37s) & & (0.44s) & & (1028.13s) & & (0.62s) & & (0.73s) & & (1.18s) \\
        \hline

        \multirow{2}{*}{MKT06}
        & 223 & \multirow{2}{*}{8\%} & 238 & \multirow{2}{*}{15\%} & 219.5 & \multirow{2}{*}{6\%} & 248 & \multirow{2}{*}{20\%} & 213.5 & \multirow{2}{*}{3\%} & \textbf{206.5} & \multirow{2}{*}{\textbf{0\%}} & 217 & \multirow{2}{*}{5\%} \\
        & \textbf{(0.51s)} & & (0.52s) & & (0.61s) & & (1013.68s) & & (0.94s) & & (1.02s) & & (1.69s) \\
        \hline

        \multirow{2}{*}{MKT07}
        & 430 & \multirow{2}{*}{24\%} & 444 & \multirow{2}{*}{28\%} & 399 & \multirow{2}{*}{15\%} & 367 & \multirow{2}{*}{5\%} & 406 & \multirow{2}{*}{17\%} & 427 & \multirow{2}{*}{23\%} & \textbf{348} & \multirow{2}{*}{\textbf{0\%}} \\
        & (0.35s) & & \textbf{(0.34s)} & & (0.41s) & & (1027.07s) & & (0.59s) & & (0.69s) & & (1.12s) \\
        \hline

        \multirow{2}{*}{MKT08}
        & 850.5 & \multirow{2}{*}{13\%} & 858.5 & \multirow{2}{*}{14\%} & 836.5 & \multirow{2}{*}{11\%} & \textbf{752} & \multirow{2}{*}{\textbf{0\%}} & 819 & \multirow{2}{*}{9\%} & 828 & \multirow{2}{*}{10\%} & 812 & \multirow{2}{*}{8\%} \\
        & \textbf{(0.79s)} & & (0.79s) & & (0.93s) & & (1053.28s) & & (1.42s) & & (1.57s) & & (2.56s) \\
        \hline

        \multirow{2}{*}{MKT09}
        & 685.5 & \multirow{2}{*}{30\%} & 704.5 & \multirow{2}{*}{33\%} & 689 & \multirow{2}{*}{30\%} & 610 & \multirow{2}{*}{16\%} & 682.5 & \multirow{2}{*}{29\%} & \textbf{528} & \multirow{2}{*}{\textbf{0\%}} & 529 & \multirow{2}{*}{0\%} \\
        & \textbf{(0.84s)} & & (0.84s) & & (1.0s) & & (1024.91s) & & (1.52s) & & (1.66s) & & (2.72s) \\
        \hline

        \multirow{2}{*}{MKT10}
        & 595 & \multirow{2}{*}{45\%} & 598 & \multirow{2}{*}{46\%} & 587.5 & \multirow{2}{*}{44\%} & 509 & \multirow{2}{*}{24\%} & 531.5 & \multirow{2}{*}{30\%} & 417 & \multirow{2}{*}{2\%} & \textbf{409} & \multirow{2}{*}{\textbf{0\%}} \\
        & (0.84s) & & \textbf{(0.83s)} & & (1.0s) & & (1036.62s) & & (1.52s) & & (1.64s) & & (2.72s) \\

        \hline
        \hline

        Average & - & 28\% & - & 35\% & - & 23\% & - & 12\% & - & 14\% & - & 10\% & - & \textbf{3\%} \\
        
        \Xhline{3\arrayrulewidth}
    \end{tabular}
    }
    \caption{Makespan and runtime results on benchmark dataset.}
    \label{tab:benchmark_dataset1}
    \vspace{-20pt}
\end{table*}

\begin{figure}[t]
    \centering
    \subfigure[10$\times$6$\times$6]{
    \includegraphics[width=0.45 \linewidth]{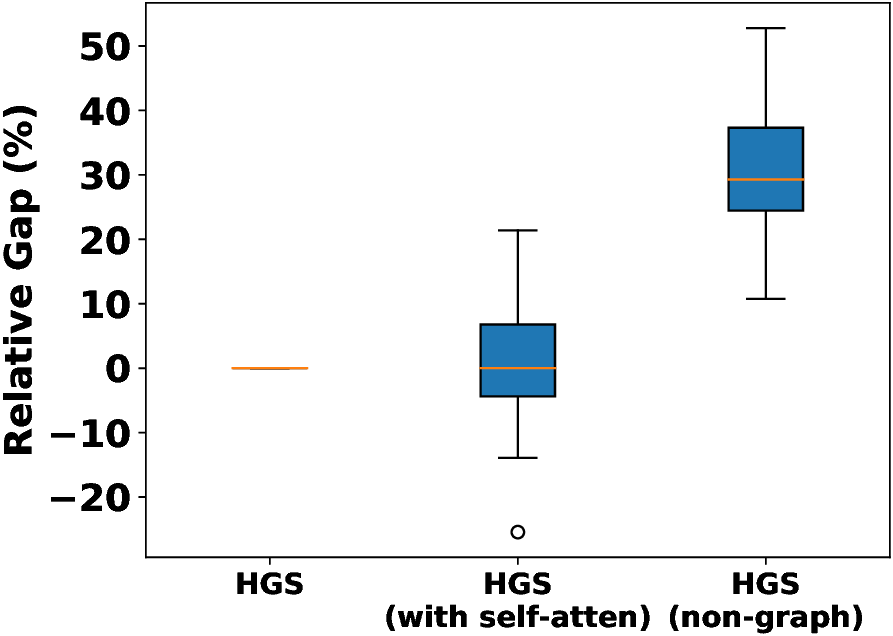}
    }
    \subfigure[50$\times$25$\times$25]{
    \includegraphics[width=0.45 \linewidth]{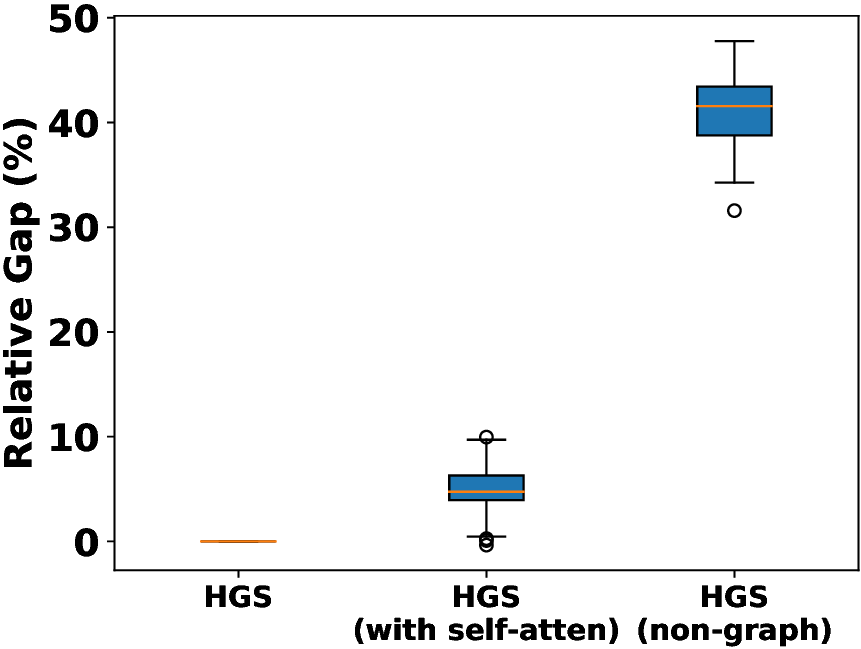}
    }
    \caption{Relative gap distribution between transformed \our{} methods on various instances.}
    \label{fig:analy_HGS}
    \vspace{-15pt}
\end{figure}

\subsection{Analysis of the proposed method}
In this section, we analyze the proposed method how to achieve the above superiority.
To this end, we compare \our{} with several transformed \our{} modules:
\begin{itemize}
    \item \our{} (non-graph): This variant is created to validate the effectiveness of the proposed encoder-decoder models. It employs a non-graph-based DRL method, using a basic attention-based decoder~\cite{b8} to form a composite action while maintaining the size-agnostic property. Here, $\pi(a_t|s_t) = \pi(O_{ij}|s_t) \pi(M_k|s_t) \pi(V_u|s_t)$, and the state $s_t$ represents the initial node embeddings without encoding them.
    \item \our{} (self-attention): This variant uses a self-attention encoder~\cite{b8} in place of the proposed encoder, but retains the proposed decoder. The self-attention encoder assumes that a node is initially related to all other nodes.
\end{itemize}
These models are trained on a graph of size 10$\times$6$\times$6 and tested on graphs 50$\times$25$\times$25. 
We conduct 100 tests for each instance and examine the distribution of the makespan gap of the transformed methods relative to \our{}. The results are depicted in Fig.~\ref{fig:analy_HGS}.

The \our{} (non-graph) variant exhibits an average gap of 29\% compared to \our{} on the 10$\times$6$\times$6 trained instance. However, this gap increases to 41\% on unseen instances of size 50$\times$25$\times$25. The significant performance difference, even on trained instances, suggests that the graph-based DRL method is more proficient at determining actions in FJSPT.
Conversely, \our{} (self-attention) performs similarly to \our{} on the trained instance, with an approximate gap of 1\%, and even produces better solutions in some tests. However, this gap increases to 5\% on unseen large-scale instances of size 50$\times$25$\times$25. In all tests, \our{} outperforms \our{} (self-attention). 
The one-to-all connection in \our{} (self-attention) proves challenging to generalize as the graph size increases, due to the complex node relationships. 
These observations suggest that the sub-graph-based one-to-few connection in \our{} offers superior generalization to unseen large-scale instances, attributable to fewer node connections and the inductive bias between node classes.

\subsection{Benchmark test} \label{sec:benchmark_test}
In this section, we demonstrate the proposed method on two FJSPT benchmark datasets referenced in paper~\cite{b25}.
This dataset, originally proposed by Brandimarte~\cite{b4}, consists of ten instances, which involve 10, 15, and 20 jobs, 55-240 operations, and 4-15 machines. 
It incorporates machine layout with transportation time between machines, where the time is randomly generated between 2 and 10.
The number of vehicles $v$ for each instance is sampled from distribution $\text{U}(0.8m, 1.2m)$.
DRL-based methods (\our{}, HGNN, MatNet) use the model trained on graph size 10$\times$6$\times$6.

As depicted in Table~\ref{tab:benchmark_dataset1}, the proposed method finds the best solutions for most instances, with the exception of only three instances (MKT01, 06 and 08).
When comparing with DRL-based methods (HGNN and MatNet), the proposed method significantly enhances performance, with the average gap difference reaching up to 9\%.
Compared with the IGA method (average gap of 12\%), \our{} yields better results and is much less computationally expensive.

\section{Conclusion}
In AGV-based SMSs, scale generalization is a challenge that DRL methods should address to maximize productivity in a variety of manufacturing environments.
In this paper, we have proposed a novel graph-based DRL method, called \our{}, to address the challenge of scale-generalizable FJSPT.
Compared to the existing dispatching rules, meta-heuristic and DRL-based methods, we have demonstrated that the proposed method provides superior solutions in terms of makespan minimization with a reasonable computation efficiency.
In particular, we have observed that the proposed heterogeneous graph encoder contributes to scale generalization even on unseen large-scale instances.


\ifCLASSOPTIONcaptionsoff
  \newpage
\fi

\bibliographystyle{IEEEtran}
\bibliography{sample-bibliography.bib}

\end{document}